\documentclass[3p, review, sort&compress]{elsarticle}

\usepackage[T1]{fontenc}

\usepackage{amssymb}
\usepackage{amsmath}

\usepackage{gensymb}
\usepackage[dvipsnames]{xcolor}
\usepackage{soul}

\usepackage[free-standing-units=true]{siunitx}
\usepackage[version=4]{mhchem}
\usepackage{csquotes}
\usepackage{stackengine}
\stackMath
\newcommand\tsup[2][2]{%
 \def\useanchorwidth{F}%
  \ifnum#1>1%
    \stackon[-1.3ex]{\tsup[\numexpr#1-1\relax]{#2}}{\kern0.2em\mathchar"307E}%
  \else%
    \stackon[-1ex]{#2}{\kern0.2em\mathchar"307E}%
  \fi%
}

\journal{Photoacoustics}

\usepackage{etoolbox}
\patchcmd{\maketitle}
  {\finalMaketitle}
  {\finalMaketitle{\includegraphics[height=\the\baselineskip]{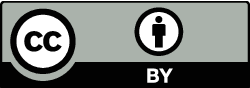} This work is licensed under the \href{http://creativecommons.org/licenses/by/4.0/}{Creative Commons Attribution 4.0 International License}.}\vspace{\baselineskip}}
  {}{}

\usepackage[hidelinks, colorlinks=true]{hyperref}

\begin{document}
\begin{frontmatter}



\title{Time-domain Brillouin scattering theory for probe light and acoustic beams propagating at an angle and acousto-optic interaction at material interfaces}


\author[inst1]{Vitalyi E. Gusev\corref{cor1}}
\ead{vitali.goussev@univ-lemans.fr}
\cortext[cor1]{Corresponding authors}

\affiliation[inst1]{organization={Laboratoire d'Acoustique de l'Universit\'e du Mans (LAUM), UMR 6613, Institut d'Acoustique -- Graduate School (IA-GS), CNRS, Le Mans Universit\'e},
            country={France}}

\author[inst1]{Th\'{e}o Thr\'{e}ard}
\author[inst2]{David H. Hurley}

\affiliation[inst2]{organization={Idaho National Laboratory},
            addressline={P.O. Box 1625}, 
            city={Idaho Falls},
            postcode={83415}, 
            state={ID},
            country={USA}}

\author[inst1]{Samuel Raetz\corref{cor1}}
\ead{samuel.raetz@univ-lemans.fr}

\begin{abstract}
A theory has been developed to interpret time-domain Brillouin scattering (TDBS) experiments involving coherent acoustic pulse (CAP) and light pulse beams propagating at an angle to each other. It predicts the influence of the directivity pattern of their acousto-optic interaction on TDBS signals when heterodyne detection of acoustically scattered light is in backward direction to incident light. The theory reveals the relationships between the carrier frequency, amplitude and duration of acoustically induced "wave packets" in light transient reflectivity signals, and factors such as CAP duration, widths of light and sound beams, and their interaction angle. It describes the transient dynamics of these wave packets when the light and CAP encounter material interfaces, and the light scattering by the incident CAP transforms into scattering by the reflected and transmitted CAPs. The theory suggests that single-point TDBS experiments can determine not only the depth positions of buried interfaces but also their inclinations/orientations.
\end{abstract}





\end{frontmatter}


\section{Introduction}
\label{sec:intro}
In picosecond acoustics, ultrafast lasers are generating and detecting coherent acoustic pulses (CAPs) \cite{thomsen_coherent_1984,thomsen_surface_1986,grahn_picosecond_1989,akhmanov_laser_1992,gusev_laser_1996,matsuda_fundamentals_2015}, which can be applied for materials and structures evaluation at sub-micrometers to nanometers spatial scale along their propagation direction. In the context of evaluation of transparent materials/structures, the technique is called either picosecond acoustic (ultrasonic) interferometry or time-domain Brillouin scattering (TDBS) \cite{grahn_picosecond_1989,thomsen_picosecond_1986,lin_phonon_1991,wright_ultrafast_1991,wright_thickness_1992,ohara_characterization_2001,devos_different_2005,hudert_influence_2008,lomonosov_nanoscale_2012,nikitin_revealing_2015,gusev_advances_2018}. The applications of picosecond acoustics are numerous, examples of which can be found in reviews~\cite{gusev_advances_2018,maris_picosecond_1998,antonelli_characterization_2006, audoin2023}, as well as in a special section of the journal Ultrasonics dedicated to that matter and published in 2015~\cite{PERONNE2015}.

From the mathematical point of view, the theory describing the scattering of plane probe light wave by plane acoustic waves composing the CAPs and propagating collinearly to probe light was sufficient for the interpretation of experimental results for many years. This is because most of the applications of the TDBS were focusing on the evaluation of laterally homogeneous materials or layered samples structured along the direction normal to the surface, such as films on substrates and superlattices. Although the role of the diffraction of light and sound beams of equal radii was discussed for the first time already in one of the pioneering publications \cite{lin_phonon_1991}, the interest for this role has only quite recently grown with TDBS experiments using diffracting (optical and/or acoustical) beams. This has happened since experimental applications of CAPs generation via femtosecond laser pulses absorption in nanometers-to-micrometers-size transducers were achieved \cite{dehoux_relaxation_2012, dehoux_optical_2016, xu_all-optical_2018}, where the dimensions of the transducers are not much larger than the acoustic wavelengths in the TDBS experiments. Current interest to the scattering of optical beam by acoustical beam is also due to the progress in TDBS applications for microscopy, where tight focusing of the pump and probe laser beams is required to achieve a micrometer-scale lateral resolution \cite{yu_brillouin_2017, devkota_attenuation_2019, perez-cota_apparent_2020}. The influence of the diffraction effects on the TDBS has been revealed in several experiments \cite{dehoux_relaxation_2012, dehoux_optical_2016, xu_all-optical_2018, yu_brillouin_2017, devkota_attenuation_2019, perez-cota_apparent_2020} and the theory for the TDBS in collinear paraxial light and sound beams has been developed \cite{gusev_contra-intuitive_2020}. Additional extensions of the TDBS theory are currently required for some recently emerged applications of the TDBS to two- and three-dimensional imaging, where the arrivals of the CAPs on inclined (not normal to the directions of sound and probe light propagation) inter-grain boundaries in polycrystalline materials were experimentally observed \cite{khafizov_subsurface_2016, wang_nondestructive_2019, wang_imaging_2020, sandeep_3d_2021, threard_photoacoustic_2021}. Very recently, the changes of the TDBS signal accompanying the CAP and the probe incidence on the inclined interface between two acoustically and optically isotropic homogeneous materials were revealed and reported \cite{la_cavera_phonon_2021}. In a transmission through or a reflection from inclined interface between grains, the angle between initially collinear propagation directions of the probe light beam and the CAP beam could change. Thus, the development of a TDBS theory for probe light and acoustic beams propagating at an arbitrary angle is needed to more accurately analyze experimental results. 

In this manuscript, a simple theory describing the dynamics of the TDBS signals when CAPs are interacting with material inclined interfaces is developed. In Sec.~\ref{sec:TDBS_arbitratry_acField}, a method of light beams decomposition into plane waves, applied recently for the analysis of their acousto-optic interaction with paraxial CAP beams \cite{gusev_contra-intuitive_2020}, is reviewed and extended to the case of arbitrary acoustic field detection via $180\degree$ backward probe light scattering, \textit{i.e.}, in the direction opposite to the incident probe light and collinear to the direction of the probe light reflected by the stationary surfaces/interfaces of the sample. The general theoretical formula suggested in Sec.~\ref{sec:TDBS_arbitratry_acField} is applied in Sec.~\ref{subsec:free_space} for the analytical description of TDBS in coherent Gaussian probe light and acoustic beams propagating at an arbitrary angle in isotropic homogeneous materials. TDBS transient signals, resulting from the acousto-optic interaction of light and sound beams in homogeneous materials far from any interfaces are predicted, and the directivity pattern of the TDBS detection via a non-diffracting probe light beam is revealed. In Secs.~\ref{subsec:reflection} and \ref{subsec:transmission}, the theoretical predictions are extended to the cases of CAP reflection and transmission, respectively, at an elastic interface corresponding to a plane boundary between two isotropic materials with different acoustical but equal optical properties. In Secs.~\ref{subsec:reflection_with_light_reflection} and \ref{subsec:transmission_with_light_reflection}, the generalization of the theory is developed for the experimental configurations where not only the CAP beam but also the probe light beam is reflected and refracted by the interface. In all experimental configurations considered in Sec.~\ref{sec:TDBS_arbitratry_angle}, the analytical results are derived under a highly relevant assumption of Gaussian pump/probe laser beams. Thus, cross sections of the probe light beam and the CAP beam, the latter replicating that of the pump light laser focus on the optoacoustic generator \cite{akhmanov_laser_1992, lin_phonon_1991, gusev_laser_1993}, are both assumed to be Gaussian. This completely analytical approach provides opportunity to describe by compact formulas all characteristic scales in time and relative propagation angle relevant to the detection of non-collinear light and sound beams scattering at the angle imposed by the heterodyning detection of the backscattered probe light. Section~\ref{sec:discussion} includes the summary and the discussion of the assumptions introduced for the development of the theory and the comments on how the theory could be extended to the case of the boundary between acoustically and optically anisotropic grains/media. The experimental situation where the inclination of the interface introduces the directions of acoustically-scattered probe light heterodyning that are additional to the backward reflection direction of the probe light is eventually discussed.

\section{TDBS of probe light by an arbitrary acoustic field}
\label{sec:TDBS_arbitratry_acField}
\begin{figure}
    \centering
    \includegraphics{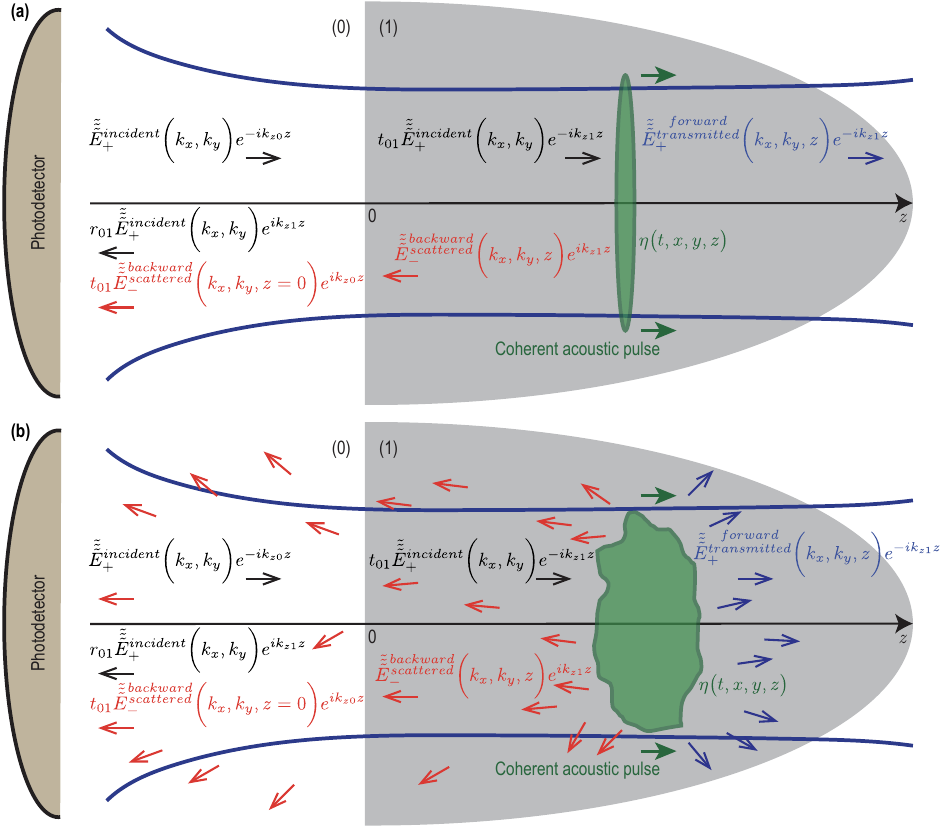}
    \caption{Schematic representation of the TDBS experiment in the common case of probe light beam normally incident on the plane surface of a semi-infinite sample. (a) A particular case of co-focused acoustic and probe light beams. The coherent acoustic pulse contributing to the time-domain Brillouin scattering signal detection is represented at a given position along its propagation path. The initial probe light field, that exists even in the absence of the coherent acoustic field, is due to the laser pulses incident on the sample surface ($z=0$) from the semi-infinite light-transparent material denoted by (0) ($z<0$). These incident pulses induce the laser pulses transmitted inside the semi-infinite material denoted by (1) ($z>0$) and the laser pulses reflected from the sample surface. The coherent acoustic (strain) pulse $\eta(t,x,y,z)$ is launched in the transparent material (1) due to the absorption, near the surface, of the pump laser beam (not represented), co-focused on the surface $z=0$ with the probe laser beam. Either the material (1) is highly absorbing for pump laser pulses or a semi-transparent, usually thin, metallic layer (not represented) is deposited on the surface. When the pump laser pulses are incident on the sample before the probe laser pulses, the acousto-optic interaction between the coherent acoustic pulses and the initial probe light field creates the optical polarization sources of the scattered light. The time-domain Brillouin scattering signal is a result of the heterodyning of the acoustically scattered probe light by the surface-reflected probe light, both collected by the same photodetector. (b): A general case of the TDBS by an arbitrary acoustic strain field in the half space. The heterodyne detection selects the probe light acoustically scattered along the direction of the reflected (heterodyning) probe light.}
    \label{fig:fig1_arb_ac_field}
\end{figure}

In Fig.~\ref{fig:fig1_arb_ac_field}(a), the most typical configuration of TDBS experiments is presented, \textit{i.e.}, the co-focused acoustic and light beams propagating along the same direction. The sample is either a strong absorber of the pump laser radiation or its surface is covered by a thin, semi-transparent metallic layer. In both cases the absorption of ultrashort pump laser pulses serves to launch picosecond CAPs in the sample. Quite recently, previous theoretical analyses, conducted in Refs. \cite{thomsen_coherent_1984, thomsen_surface_1986,grahn_picosecond_1989} for the collinear propagation of plane probe light and CAPs and in Ref. \cite{lin_phonon_1991} for diffracting light and sound beams of equal initial radii, have been extended (generalized) to the case of the collinear propagation of a diffracting CAP beam and a diffracting probe light beam of different cross sections \cite{gusev_contra-intuitive_2020}.

Note that, in Fig.~\ref{fig:fig1_arb_ac_field}(a), the acoustic strain field is presented as a CAP of finite spatial dimensions propagating in a particular direction normal to the sample surface just to refer to the configuration which is the most common to the experiments and the earlier developed theories \cite{lin_phonon_1991,gusev_contra-intuitive_2020}, while the theory developed in this Section addresses an arbitrary strain field in the half space [Fig.~\ref{fig:fig1_arb_ac_field}(b)], including the particular cases depicted in Fig.~\ref{fig:fig1_arb_ac_field}(a) and Fig.~\ref{fig:fig2_TDBS_freespace}.

Figure~\ref{fig:fig1_arb_ac_field} schematically presents the probe light laser beam focused through the transparent material (denoted by (0) in the half space $z<0$) on the surface ($z=0$) of the transparent sample (denoted by (1) in the half space $z>0$). It is assumed in the here-developed theory, for compactness, that the coherence length of the probe laser pulses \cite{lin_phonon_1991} significantly exceeds the depth of CAP penetration of interest. Therefore, the probe light pulses scattered by the CAP and reflected by stationary interfaces broadly overlap temporally on the photodetector. Hence, the wave-packet nature of the probe light pulses can be neglected and the probe laser radiation can be described as an unmodulated one at a single optical frequency. A superposition of the incident, transmitted and reflected probe light beams composes the initial light field distribution in the absence of the CAPs. All light beams at optical frequency $\omega$ satisfy the Helmholtz equation for the electrical field component of the laser radiation ($E(t,x,y,z)\equiv\tilde{E}(x,y,z) e^{i\omega t}$),
\begin{align}\label{eq:wave_eq}
    \left[\left(\frac{\partial^2}{\partial x^2}+\frac{\partial^2}{\partial y^2}+\frac{\partial^2}{\partial z^2}\right)+k^2\right] \tilde{E}(x,y,z)=0\,,
\end{align}
where $k(\omega)\equiv \omega/c$ denotes the optical wave number and $c$ the speed of light in the material. The solutions of Eq.~\eqref{eq:wave_eq} can be obtained via decomposition of the three-dimensional wave fields into propagating plane waves $\tsup[3]{E}(k_x,k_y,z)$, with the wave vector components $k_x$, $k_y$ and $k_z=\sqrt{k^2-k_x^2-k_y^2}$, ($\tilde{E}(x,y,z)=\iint_{-\infty}^{+\infty}\tsup[3]{E}(k_x,k_y,z)e^{-i(k_x x+k_y y)}\mathrm{d}k_x\mathrm{d}k_y$), satisfying the equation:
\begin{align}\label{eq:eq2}
    \left[\frac{\partial^2}{\partial z^2}+k_z^2\right] \tsup[3]{E}(k_x,k_y,z)=0\,.
\end{align}

The transverse distributions of the plane wave amplitudes of the transmitted and reflected light in Fig.~\ref{fig:fig1_arb_ac_field} are related at $z=0$ to the transverse distribution of the incident probe beam $\tsup[3]{E}{\vphantom{E}}^\text{incident}_+(k_x,k_y)$ via \[\tsup[3]{E}{\vphantom{E}}^\text{transmitted}_+(k_x,k_y) = t_{01}(k_x,k_y)\tsup[3]{E}{\vphantom{E}}^\text{incident}_+(k_x,k_y)\] and \[\tsup[3]{E}{\vphantom{E}}^\text{reflected}_-(k_x,k_y) = r_{01}(k_x,k_y)\tsup[3]{E}{\vphantom{E}}^\text{incident}_+(k_x,k_y)\,,\] where the transmission and reflection coefficients of the incident probe light at the surface $z=0$ are denoted by $t_{01}$ and $r_{01}$, respectively. The solution of Eq.~\eqref{eq:eq2} for the initial probe light field in Fig.~\ref{fig:fig1_arb_ac_field} is:
\begin{align}
\left\{
\begin{aligned}
    &\tsup[3]{E}{\vphantom{E}}^\text{incident}_+(k_x,k_y, z) = \tsup[3]{E}{\vphantom{E}}^\text{incident}_+(k_x,k_y)e^{-i k_{z0}z}\,,\\
    &\tsup[3]{E}{\vphantom{E}}^\text{transmitted}_+(k_x,k_y, z) = t_{01}\tsup[3]{E}{\vphantom{E}}^\text{incident}_+(k_x,k_y)e^{-i k_{z1}z}\,,\\
    &\tsup[3]{E}{\vphantom{E}}^\text{reflected}_-(k_x,k_y, z) = r_{01}\tsup[3]{E}{\vphantom{E}}^\text{incident}_+(k_x,k_y)e^{+i k_{z0}z}\,.
\end{aligned}
\right.
\end{align}
Here the indices \enquote{0} and \enquote{1} indicate the difference in the probe light velocities in the surrounding medium which is typically air (0), and the material of interest (1). When, in addition to the ultrashort probe laser pulse, the ultrashort pump laser pulse is focused on the pump-light absorbing sample surface ($z=0$), the generation of coherent acoustic waves via optoacoustic conversion takes place \cite{thomsen_coherent_1984,thomsen_surface_1986,grahn_picosecond_1989,akhmanov_laser_1992,gusev_laser_1993,ruello_physical_2015}. In Fig.~\ref{fig:fig1_arb_ac_field}(a), the diffracting strain CAP $\eta(t,x,y,z)$ launched in the material ($z>0$) is symbolically sketched for the case where the initial acoustic beam radius, controlled by the pump laser focusing, is larger than the radius of the probe light focus. The presence of the CAP in material (1) induces, via acousto-optic interaction (photoelastic effect) \cite{fabelinskii_molecular_1968, dil_brillouin_1982}, the nonlinear optical polarization \cite{xu_acousto-optic_1992}, and modifies Eq.~\eqref{eq:wave_eq}:
\begin{align}\label{eq:wave_eq_with_CAP}
    \left[\left(\frac{\partial^2}{\partial x^2}+\frac{\partial^2}{\partial y^2}+\frac{\partial^2}{\partial z^2}\right)+k_1^2\right] \tilde{E}(t,x,y,z)=k_1^2n_1^2p_1\eta(t,x,y,z)\tilde{E}(t,x,y,z)\,.
\end{align}
In Eq.~\ref{eq:wave_eq_with_CAP}, $n_1$ and $p_1$ denote the refractive index and the photoelastic constant of the material (1), the latter defined as the derivative over the strain of the inverse of the dielectric constant (permittivity), while \(n_1^2 p_1 \eta(t,x,y,z)\tilde{E}(t,x,y,z)\equiv\tilde{P}(t,x,y,z)\) is the acoustically-induced nonlinear optical polarization \cite{xu_acousto-optic_1992}. Although the time variable $t$ is back into Eq.~\eqref{eq:wave_eq_with_CAP} compared to the Helmholtz equation in Eq.~\eqref{eq:wave_eq}, it is important to bear in mind that this time dependence of the electric field $\tilde{E}$ is introduced by the presence of a propagating CAP. Due to the significantly higher velocity of light compared to acoustic velocity and the much higher frequencies of light compared to acoustic frequencies, each position of the CAP during its propagation is seen as a quasi-static strain distribution by the probe electric field. As a result, the equation describing the propagation of $\tilde{E}$ maintains the form of a Helmholtz equation, while also incorporating the acoustically-induced nonlinear optical polarization. This introduces a time-dependent parametrisation of the light field. It is important to note that this equation holds true for every moment $t$ during the propagation of the CAP. Since the acousto-optic interaction is weak, Eq.~\eqref{eq:wave_eq_with_CAP} can be solved in the single-scattering approximation \cite{thomsen_coherent_1984, ishimaru_front_1978}, where the initial \enquote{strong} light field, $\tsup[3]{E}{\vphantom{E}}^\text{transmitted}_+(k_x,k_y, z)$, creates nonlinear polarization that emits additional much weaker light waves in the material:
\begin{align}\label{eq:eq5}
\begin{split}
    \left[\left(\frac{\partial^2}{\partial x^2}+\frac{\partial^2}{\partial y^2}+\frac{\partial^2}{\partial z^2}\right)+k_1^2\right] \tilde{E}^\text{scattered}(t,x,y,z)&=k_1^2\tilde{P}^\text{initial}(t,x,y,z)\\&\equiv k_1^2n_1^2p_1\eta(t,x,y,z)\tilde{E}^\text{transmitted}(t,x,y,z)\,.
\end{split}
\end{align}

The light waves scattered by the CAP are symbolically represented by left-pointing red arrows in Fig.~\ref{fig:fig1_arb_ac_field}, both just after the scattering inside the material (1) and after the transmission from material (1) to material (0). The equation for the transverse distribution of the scattered light, derived from Eq.~\eqref{eq:eq5}, is:
\begin{align}\label{eq:eq6}
\begin{split}
    \left[\frac{\partial^2}{\partial z^2}+k_{z1}^2\right] \tsup[3]{E}{\vphantom{E}}^\text{scattered}(t,k_x,k_y,z)&=k_1^2\iint_{-\infty}^{+\infty}\tilde{P}^\text{initial}(t,x,y,z)\mathrm{d}x\mathrm{d}y\\&\equiv k_1^2\tsup[3]{P}{\vphantom{P}}^\text{initial}(t,k_x,k_y,z)\,.
\end{split}
\end{align}
The solution of Eq.~\eqref{eq:eq6} for the angular amplitude of the scattered light propagating in the negative direction of the $z$-axis at the surface $z=0$ is:
\begin{align}\label{eq:eq7}
    \tsup[3]{E}{\vphantom{E}}^\text{scattered}_-(t,k_x,k_y)=\frac{1}{2i k_{z1}}k_1^2\int_0^{+\infty}\tsup[3]{P}{\vphantom{P}}^\text{initial}(t,k_x,k_y,z^\prime)e^{-i k_{z1}z^\prime}\mathrm{d}z^\prime\,.
\end{align}
In the derivation of Eq.\eqref{eq:eq7}, it was assumed that the acoustic field is completely localized in the material ($z\ge 0$) and hence $\tilde{P}^\text{initial}(t,x,y,z<0)=0$. It is also assumed that the acoustic field is localized at finite distances near the surface, \textit{i.e.}, at the time of observation when there is no acoustic field at infinitely large distances from the surface. These two conditions provide the explanation for the integration limits over the spatial coordinate in Eq.~\eqref{eq:eq7} \cite{gusev_contra-intuitive_2020}.

At this point, it is worth reviewing the fundamentals of TDBS signal formation in picosecond acoustics. The scattered light, in Eq.~\eqref{eq:eq7}, transmitted from the material (1) to the surrounding material (0) is of much smaller amplitude than the reflected light. The information carried by the scattered light is revealed via optical heterodyning detection, through the mixing of the acoustically scattered light with the probe light reflected by the surface ($z=0$) at the photodetector, whose response is linearly proportional to the absorbed light energy and hence is quadratic in the light field amplitude. Since the photodetector collects all the light from the acoustically scattered and reflected beams, the energy at the photodetector is found from the angular distributions of the scattered and reflected light as follows:
\begin{align}\label{eq:eq8}
    \begin{split}
        W^\text{detected} &= W^\text{reflected} + W^\text{heterodyned} + W^\text{scattered}\\
        &\sim \iint_{-\infty}^{+\infty}r_{01}\tsup[3]{E}{\vphantom{E}}^\text{incident}_+(k_x, k_y)\left[r_{01}\tsup[3]{E}{\vphantom{E}}^\text{incident}_+(k_x, k_y)\right]^\ast\mathrm{d}k_x\mathrm{d}k_y\\
        &+ 2\Re\left\{\iint_{-\infty}^{+\infty}t_{10}\tsup[3]{E}{\vphantom{E}}^\text{scattered}_-(k_x, k_y)\left[r_{01}\tsup[3]{E}{\vphantom{E}}^\text{incident}_+(k_x, k_y)\right]^\ast\mathrm{d}k_x\mathrm{d}k_y\right\}\\
        &+ \iint_{-\infty}^{+\infty}t_{01}\tsup[3]{E}{\vphantom{E}}^\text{scattered}_-(k_x, k_y)\left[t_{01}\tsup[3]{E}{\vphantom{E}}^\text{scattered}_-(k_x, k_y)\right]^\ast\mathrm{d}k_x\mathrm{d}k_y
    \end{split}
\end{align}
In Eq.~\eqref{eq:eq8} and later, the symbol \enquote{$^\ast$} denotes the complex conjugation. The parameter $t_{10}$ stands for the transmission coefficient from material (1) to material (0). It is the second term of Eq.~\eqref{eq:eq8} that is relevant to the time-domain Brillouin scattering technique, as it provides the frequency mixing of the reflected and scattered light. The change in the transient optical reflectivity, \textit{i.e.}, the so-called TDBS reflectivity signal, is given by $\frac{\mathrm{d}R}{R}\equiv W^\text{heterodyned}/W^\text{reflected}$. The most important feature here is that heterodyning detection with weakly diffracting probe radiation precisely selects, from the total scattered light field, the components that are propagating quasi-collinearly to the probe light beam axis: the heterodyning detection is very directive \cite{gusev_contra-intuitive_2020}.
When the plane wave decomposition of the transmitted light field,
\[
    \tilde{E}^\text{transmitted}(x,y,z) = \frac{1}{(2\pi)^2}\iint_{-\infty}^{+\infty}t_{01}\tsup[3]{E}{\vphantom{E}}^\text{incident}_+(k_x^\prime, k_y^\prime)e^{-i\left(k_x^\prime x + k_y^\prime y\right)}e^{-i k_{z1}^\prime z}\mathrm{d}k_x^\prime\mathrm{d}k_y^\prime\,,
\]
with $k_{z1}^\prime=\sqrt{k_1^2-k_x^{\prime 2}-k_y^{\prime 2}}$, is substituted in the polarization $\tilde{P}^\text{initial}(t,x,y,z)$ [Eq.~\eqref{eq:eq5}], the scattered field can be evaluated with Eq.~\eqref{eq:eq7} and, finally, a rather general presentation of the heterodyned contribution to the light energy on the photodetector is derived:
\begin{align}\label{eq:eq9}
\begin{split}
    W^\text{heterodyned} \sim -\frac{k_1^2n_1^2p_1}{(2\pi)^2}\Im\left\{\iint_{-\infty}^{+\infty}\left[r_{01}\tsup[3]{E}{\vphantom{E}}^\text{incident}_+(k_x, k_y)\right]^\ast\frac{t_{10}}{k_{1z}}\left[\iint_{-\infty}^{+\infty}\left(\int_0^{+\infty}\eta(t,x,y,z)\iint_{-\infty}^{+\infty}t_{01}\right.\right.\right.\\\left.\left.\left.\times\tsup[3]{E}{\vphantom{E}}^\text{incident}_+(k_x^\prime, k_y^\prime)e^{i\left(k_x-k_x^\prime\right)x + i\left(k_y-k_y^\prime\right)y}e^{-i \left(k_{z1} + k_{z1}^\prime\right) z}\mathrm{d}k_x^\prime\mathrm{d}k_y^\prime\mathrm{d}z\vphantom{\iint_{-\infty}^{+\infty}}\right)\mathrm{d}x\mathrm{d}y\right]\mathrm{d}k_x\mathrm{d}k_y\right\}\,.
\end{split}
\end{align}
The only assumptions to get the solution in Eq.~\eqref{eq:eq9} are the perfect transparency of the material (1) at the wavelength of the probe light, leading to real-valued $k_1$, $n_1$ and $p_1$, and the opportunity to describe the acousto-optic interaction by a single constant $p_1$ (see Sec.~\ref{sec:discussion} for a discussion on some of the introduced assumptions).

Recently, Eq.~\eqref{eq:eq9} has been applied to analyze the TDBS signal in the case of weakly diffracting probe light field \cite{gusev_contra-intuitive_2020}. In the paraxial approximation of the diffraction theory \cite{akhmanov_physical_1997}, the projections of the entire light wave vectors on the $z$ direction are approximated, in the description of the wave amplitude, by the full wave vectors, \textit{i.e.}, the transverse wave vector components are neglected. Simultaneously, the possible dependences of all transmission/reflection coefficients on the transverse wave vector components are neglected and they are treated as constants, for example $t_{01}(k_x,k_y)\cong t_{01}(0,0)\equiv t_{01}$ and $r_{01}(k_x,k_y)\cong r_{01}(0,0)\equiv r_{01}$. In the description of the phases, the first order corrections, proportional to the square of the transverse-to-axial component ratio, is taken into account. Under the listed assumptions, Eq.~\eqref{eq:eq9} takes the following simplified form:
\begin{multline}\label{eq:eq10}
    W^\text{heterodyned} \sim -\frac{k_1n_1^2p_1}{(2\pi)^2}r_{01}\left(1-r_{01}^2\right)\Im\left\{\iint_{-\infty}^{+\infty}\int_0^{+\infty}\eta(t,x,y,z)e^{-i 2k_1z}\left(\iint_{-\infty}^{+\infty}\left[\tsup[3]{E}{\vphantom{E}}^\text{incident}_+(k_x, k_y)\right]^\ast\right.\right.\\\left.\left.\times\iint_{-\infty}^{+\infty}\tsup[3]{E}{\vphantom{E}}^\text{incident}_+(k_x^\prime, k_y^\prime)e^{i\left(k_x-k_x^\prime\right)x + i\left(k_y-k_y^\prime\right)y}\right.\right.\\\left.\left.\times e^{\frac{i}{2k_1}\left(k_x^2+k_y^2+k_x^{\prime 2}+k_y^{\prime 2}\right) z}\,\mathrm{d}k_x^\prime\mathrm{d}k_y^\prime\,\mathrm{d}k_x\mathrm{d}k_y\vphantom{\iint_{-\infty}^{+\infty}}\right)\mathrm{d}z\mathrm{d}x\mathrm{d}y\right\}\,.
\end{multline}

The theoretical result in Eq.~\eqref{eq:eq10} has been applied in \cite{gusev_contra-intuitive_2020} to investigate the experimental situations where the coherent acoustic strain field $\eta(t,x,y,z)$ is itself a paraxial acoustic beam oriented along the same axis as the probe laser beams. It has been revealed that, counterintuitively, the TDBS signal amplitude variations with time do not depend on the variations of the paraxial CAP amplitude in the sound beam diffraction process. The key origin of this phenomenon is the phase-sensitive process of the acoustically scattered and reflected probe light interference in the heterodyning detection. We would like to emphasis hither that the theoretical solution in Eq.~\eqref{eq:eq10} could be very useful for the analytical interpretation of various other experimental configurations. A particular case of interest that we discuss hereafter is the TDBS configuration where the probe laser beam is non diffracting. Omitting the phase factor $\frac{i}{2k_1}\left(k_x^2+k_y^2+k_x^{\prime 2}+k_y^{\prime 2}\right)z$ responsible for the diffraction of the paraxial light beam in Eq.~\eqref{eq:eq10}, the description of the TDBS signal takes an extremely insightful form:
\begin{align}\label{eq:eq11}
\begin{split}
    W^\text{heterodyned} &\sim -\frac{k_1n_1^2p_1}{(2\pi)^2}r_{01}\left(1-r_{01}^2\right)\Im\left\{\iint_{-\infty}^{+\infty}\int_0^{+\infty}\eta(t,x,y,z)\tilde{E}^\text{incident}_+(x, y, 0)\right.
    \\
    &\qquad\qquad\qquad\qquad\qquad\qquad\qquad\qquad\qquad~\times\left.\left[\tilde{E}^\text{incident}_+(x, y, 0)\right]^\ast e^{-i 2k_1z}\,\mathrm{d}z\mathrm{d}x\mathrm{d}y\vphantom{\iint_{-\infty}^{+\infty}}\right\}\,,
    \\
    \vphantom{W^\text{heterodyned}} &\sim +\frac{k_1n_1^2p_1}{(2\pi)^2}r_{01}\left(1-r_{01}^2\right)\iint_{-\infty}^{+\infty}\int_0^{+\infty}\eta(t,x,y,z)I_+^\text{incident}(x,y,0) \sin\left(2k_1z\right)\mathrm{d}z\mathrm{d}x\mathrm{d}y\,.
\end{split}
\end{align}
In accordance with Eq.~\eqref{eq:eq11}, the TDBS signal is related to the $2k_1$ spatial Fourier component, along the propagation direction of the probe light, of the three-dimensional spatial overlap between the cylindrical probe light intensity column, $I_+^\text{incident}(x,y,0)$, and the 3D acoustic strain field in half-space or, said differently, to the Laplace transform of complex parameter \cite{gusev_laser_1996}. In the classical configuration of a plane probe light pulse and a plane CAP both propagating along the $z$-axis, Eq.~\eqref{eq:eq11} reproduces the well-known result of the one-dimensional theory:
\begin{align}\label{eq:eq12}
    W^\text{heterodyned} \sim -\frac{k_1n_1^2p_1}{(2\pi)^2}r_{01}\left(1-r_{01}^2\right)I_+^\text{incident}(z=0)\int_{0}^{+\infty}\eta(t,z)\sin\left(2k_1z\right)\,\mathrm{d}z\,.
\end{align}
Equation~\eqref{eq:eq11} could of course be extended to the case where the material (1) is absorbing the probe light by making use of its complex refractive index and its complex photo-elastic constant. The most important point for our subsequent analysis is the fact that Eq.~\eqref{eq:eq11} is not only valid for plane probe laser fields but also for arbitrary probe light beams inside their Rayleigh range (at distances shorter than the diffraction length) \cite{akhmanov_physical_1997, siegman_introduction_1968,self_focusing_1983}. At distances larger than the diffraction length from the surface of the material (1), the omitted phase factor $\frac{i}{2k_1}\left(k_x^2+k_y^2+k_x^{\prime 2}+k_y^{\prime 2}\right)z$, accumulating with depth, does not lead to non-negligible diminishing of the probe light field amplitude. Inside the Rayleigh range of the probe light field, the theoretical formula Eq.~\eqref{eq:eq11} is valid for an arbitrary transient acoustic strain field localized inside the material (1).

In summary, the analytical calculation described in Section~\ref{sec:TDBS_arbitratry_acField} has resulted in a theoretical formula that predicts the TDBS signal for an arbitrary acoustic field in the most commonly used experimental geometry that involves the propagation of the probe incident and heterodyning light beams in opposite directions. The formula represents a generalization of the previously known theory, which was based on collinear propagation of the plane probe light and CAP. However, the new formula does not rely on the assumption of collinearity and accommodates non-diffracting probe light and CAP beams.

\section{TDBS in coherent probe light and acoustic beams propagating at an arbitrary angle}
\label{sec:TDBS_arbitratry_angle}
Starting from the pioneering work in picosecond laser ultrasonics \cite{thomsen_coherent_1984, thomsen_surface_1986, grahn_picosecond_1989} up to very recent times, the theoretical descriptions obtained for collinear (anti-collinear) plane probe light and CAP propagation of the type of Eq.~\eqref{eq:eq12}, with the extension to the structures layered along the $z$-axis (refer to Ref.~\cite{gusev_laser_1996}, for example), were mostly sufficient to interpret experimental data and to extract the parameters of the studied structure (material (1) alone or multilayers). A handful of experimental geometries require to take into account not only backward Brillouin scattering, as in Sec.~\ref{sec:TDBS_arbitratry_acField}, but also forward Brillouin scattering \cite{matsuda_time-domain_2018, matsuda_optical_2020}. However, recent experiments on TDBS imaging of transparent polycrystalline materials \cite{khafizov_subsurface_2016, wang_imaging_2020, sandeep_3d_2021, threard_photoacoustic_2021} revealed the refraction of CAPs at sub-surface grain boundaries, accompanied by characteristic changes in the TDBS signal amplitude and/or frequency spectrum. Since the CAP transmission through, and reflection from, a material interface inclined with respect to the CAP propagation direction is accompanied by changes in the CAP propagation direction \cite{wang_imaging_2020, sandeep_3d_2021, threard_photoacoustic_2021, la_cavera_phonon_2021}, the development of the TDBS theory for probe light and sound beams propagating at an angle is required. In the following sub-sections, it is demonstrated that the description of the temporal dynamics of the TDBS signal modifications, accompanying the CAP interaction with the inclined material interface, can be obtained through a dedicated analysis of the theoretical solution derived in Eq.~\eqref{eq:eq11}. In all the depicted geometrical configurations, including probe light and CAP beam interactions in homogeneous material or near inclined material interfaces, the analytical results are derived under a highly relevant assumption of lateral Gaussian intensity distribution of the pump and probe laser beams \cite{self_focusing_1983}. In all the configurations analyzed in the remaining part of the manuscript dealing with an inclined interface between two isotropic homogeneous materials, the assumption is made that CAPs are generated by the Gaussian pump laser beam absorption at the surface of the first material (1), that is to say that the acoustic generation in the second material (2) is neglected (see Sec.~\ref{sec:discussion} for a discussion on the latter assumption).

\subsection{Non-collinear interaction of acoustic and light beams in free space}
\label{subsec:free_space}
The analytical formulas established in Sec.~\ref{sec:TDBS_arbitratry_acField} are applied to develop the TDBS signal of interest:
\begin{align}\label{eq:eq13}
    \frac{\mathrm{d}R}{R}\equiv\frac{W^\text{heterodyned}}{W^\text{reflected}}\equiv B=P\frac{\iint_{-\infty}^{+\infty}\int_0^{+\infty}\eta(t,x,y,z)I_+^\text{incident}(x,y,0) \sin\left(2k_1z\right)\mathrm{d}z\mathrm{d}x\mathrm{d}y}{\iint_{-\infty}^{+\infty}I_+^\text{incident}(x,y,0)\,\mathrm{d}x\mathrm{d}y}\,,
\end{align}
where the following compact notation is introduced: $P\equiv k_1n_1^2p_1\left(1-r_{01}^2\right)/r_{01}$.

The Gaussian distribution of the probe light field at $z=0$, \textit{i.e.}, at the focus, is defined as $E_+^\text{incident}(x,y,0)=E_0\exp\left[-(x^2+y^2)/a_\text{probe}^2\right]$, where $a_\text{probe}$ denotes the radius of the intensity distribution at $1/e^2$ level. Equation~\eqref{eq:eq13} hence takes the form:
\begin{align}\label{eq:eq14}
    \frac{\mathrm{d}R}{R}\equiv B=\frac{2}{\pi a_\text{probe}^2}P\iint_{-\infty}^{+\infty}\exp\left(-2\frac{x^2+y^2}{a_\text{probe}^2}\right)\left[\int_0^{+\infty}\eta(t,x,y,z)\sin\left(2k_1z\right)\mathrm{d}z\right]\mathrm{d}x\mathrm{d}y\,.
\end{align}

\begin{figure}
    \centering
    \includegraphics{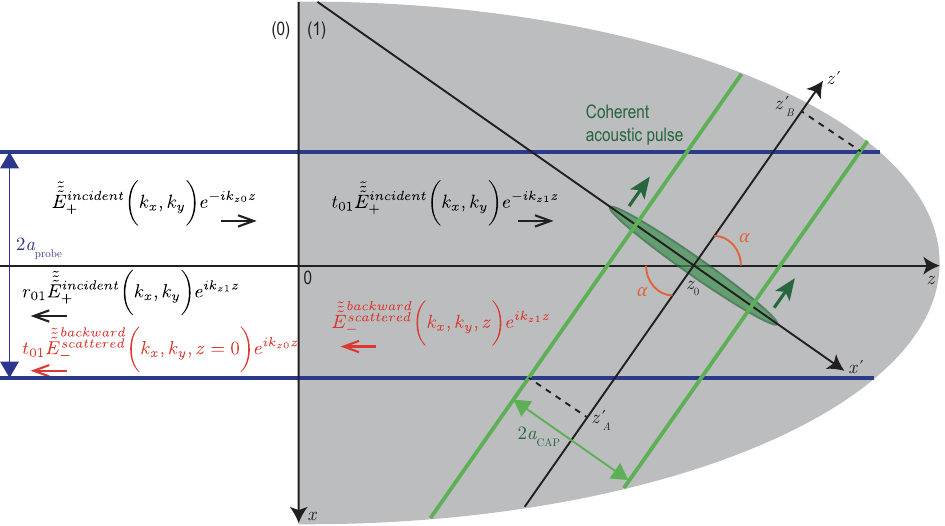}
    \caption{TDBS monitoring by the probe light beam (radius $a_\text{probe}$) of the CAP beam (radius $a_\text{\tiny CAP}$) propagating at an angle $\alpha$ relative to the probe light path direction. The axes of these two Gaussian beams are in the figure plane and intersect at the point of coordinates $(x=0,y=0,z=z_0)$.}
    \label{fig:fig2_TDBS_freespace}
\end{figure}

In the upcoming developments, Eq.~\eqref{eq:eq14} is applied to the case of a coherent acoustic strain field $\eta(t,x,y,z)$ taking the form of a directional non-diffracting Gaussian beam. We apply here the formula [Eq.~\eqref{eq:eq14}] for the analysis of the TDBS signal in the case of a coherent acoustic beam (CAP beam) propagating in a tilted direction relative to that of the probe light path, as illustrated schematically in Fig.~\ref{fig:fig2_TDBS_freespace}.

Figure~\ref{fig:fig2_TDBS_freespace} introduces the propagation of both the light and sound beams in the $y=0$ plane. The CAP beam propagates along the $z^\prime$ direction, inclined at an angle $\alpha$ to the $z$ direction, which is that of the probe light path. In the $\left(x^\prime,y^\prime,z^\prime\right)$ and the $(x,y,z)$ coordinate systems, the descriptions of the CAP beam are
\[\eta\left(t,x^\prime,y^\prime,z^\prime\right) = \eta_0\exp\left[-2\frac{x^{\prime 2}+y^{\prime 2}}{a_\text{\tiny CAP}^2}-2\frac{\left(z^\prime-v_1t\right)^2}{l_\text{\tiny CAP}^2}\right]\,,\]
and
\[\eta\left(t,x,y,z\right) = \eta_0\exp\left\{-2\frac{\left[x\cos\alpha+(z-z_0)\sin\alpha\right]^2+y^2}{a_\text{\tiny CAP}^2}-2\frac{\left[(z-z_0)\cos\alpha-x\sin\alpha-v_1t\right]^2}{l_\text{\tiny CAP}^2}\right\}\,,\]
respectively. Here $a_\text{\tiny CAP}$  and $l_\text{\tiny CAP}$ denote the radius of the Gaussian CAP beam and the half-length of the unipolar Gaussian strain pulse, both at $1/e^2$ level, respectively. $\eta_0$ is the magnitude of the strain pulse, $v_1$ stands for the acoustic wave velocity in the material (1), while $z_0$ denotes the $z$ coordinate of the intersection of the central rays of the probe light beam and the CAP beam.

The integration in Eq.~\eqref{eq:eq14} is straightforward \cite{prudnikov_integrals_1986}:
\begin{align}\label{eq:eq15}
\begin{split}
    \frac{\mathrm{d}R}{R} \equiv B
    &=\sqrt{\frac{\pi}{2}}P\left[\frac{\eta_0a_\text{\tiny CAP}^2l_\text{\tiny CAP}}{\sqrt{a_\text{probe}^2+a_\text{\tiny CAP}^2}\sqrt{a_\text{probe}^2+a_\text{\tiny CAP}^2\cos^2\alpha+l_\text{\tiny CAP}^2\sin^2\alpha}}\right.
    \\
    &\qquad\qquad\qquad\qquad\qquad\times\left.\exp\left(-\frac{k_1^2}{2}\frac{a_\text{\tiny CAP}^2l_\text{\tiny CAP}^2+a_\text{probe}^2l_\text{\tiny CAP}^2\cos^2\alpha+a_\text{probe}^2a_\text{\tiny CAP}^2\sin^2\alpha}{a_\text{probe}^2+a_\text{\tiny CAP}^2\cos^2\alpha+l_\text{\tiny CAP}^2\sin^2\alpha}\right)\right]
    \\
    &\qquad\times \exp\left[-2\frac{\sin^2\alpha}{a_\text{probe}^2+a_\text{\tiny CAP}^2\cos^2\alpha+l_\text{\tiny CAP}^2\sin^2\alpha}\left(v_1t\right)^2\right]
    \\
    &\qquad\times \sin\left\{\Omega_B(0)\left[\frac{\left(a_\text{probe}^2+a_\text{\tiny CAP}^2\right)\cos\alpha}{a_\text{probe}^2+a_\text{\tiny CAP}^2\cos^2\alpha+l_\text{\tiny CAP}^2\sin^2\alpha}t+\frac{z_0}{v_1}\right]\right\}\,,
\end{split}
\end{align}
where $\Omega_B(0)=2k_1v_1$. Eq.~\eqref{eq:eq15} predicts the TDBS signal to take a wave packet form, its envelop being described by $\exp\left(-2 t^2/\tau_B^2\right)$. The duration $\tau_B$ of the TDBS wave packet is equal to the time of the CAP propagation at velocity $v_1$ across the characteristic dimension $l_B$ of the light and sound fields overlap volume along the $z^\prime$ direction, $\tau_B=l_B/v_1$, with:
\begin{align}\label{eq:eq16}
    l_B=\sqrt{\left(\frac{a_\text{probe}}{\sin\alpha}\right)^2+\left(\frac{a_\text{\tiny CAP}}{\tan\alpha}\right)^2+l_\text{\tiny CAP}^2}\,.
\end{align}
The third term under the square root is related to the half-length of the acoustic pulse, while in the first two terms the contribution of the probe light beam and the CAP beam cross sections to the projection of the interaction volume on the $z^\prime$-axis can be revealed (Fig.~\ref{fig:fig2_TDBS_freespace}):
\[
    \frac{1}{2}\left(z_B^\prime-z_A^\prime\right)=\frac{a_\text{probe}}{\sin\alpha}+\frac{a_\text{\tiny CAP}}{\tan\alpha}\,.
\]
Equation~\eqref{eq:eq16} describes quantitatively an intuitively expected growth of the acousto-optic interaction region with all introduced characteristic spatial scales and diminishing angle of the interaction.

The monochromatic sinusoidal carrier in the TDBS wave packet, Eq.~\eqref{eq:eq15}, oscillates at the frequency \[\Omega_B=\Omega_B(0)\frac{\left(a_\text{probe}^2+a_\text{\tiny CAP}^2\right)\cos\alpha}{a_\text{probe}^2+a_\text{\tiny CAP}^2\cos^2\alpha+l_\text{\tiny CAP}^2\sin^2\alpha}=\Omega_B(0)\frac{\left(a_\text{probe}^2+a_\text{\tiny CAP}^2\right)\cos\alpha}{l_\text{B}^2\sin^2\alpha}\,,\] deviating from the so-called $180\degree$ backward-scattering Brillouin frequency case, in which $\alpha=0$ and the incident probe light wave and the acoustic wave are plane and collinearly propagate, \textit{i.e.}, $\Omega_B(0)\equiv\Omega_B(\alpha=0)=2k_1v_1$. Given plane probe light wave and plane CAP ($a_\text{probe}=+\infty$, $a_\text{\tiny CAP}=+\infty$), the carrier frequency $\Omega_B$ is equal to the Brillouin frequency shift that could be expected from the momentum conservation law in acousto-optic interaction in backward non-collinear light scattering, \textit{i.e.}, $\Omega_B\equiv\Omega_B(0)\cos\alpha$ \cite{lomonosov_nanoscale_2012, cote_refractive_2005}. For the probe light and CAP beams of finite cross sections, the carrier frequency is different from this Brillouin frequency, as it is revealed by the above-presented theoretical developments. The reason of this difference (in the geometry considered here) is found in two facts: (i) the plane acoustic and light waves that superpose to compose the corresponding beams are propagating in a variety of directions and (ii) the heterodyning detection is potentially sensitive to light scattered in the variety of directions in which the plane waves composing the incident probe light beam are reflected. Thus, the revealed carrier frequency is a result of a weighted averaging over all possible interactions between those acoustic and light plane waves providing gatherable light for the photodetector. The widths of the light and sound plane waves distributions in the wave vector space, as well as the directivity of the heterodyning detection, are all controlled by the radii of the respective beams in real space. Thus, the carrier frequency resulting from the averaging over the various interactions in the wave vector space naturally depends on the light and sound beams radii.

The presented theoretical developments, considering probe light and CAP beams of finite cross sections, reveal a shift in the carrier frequency that is different compared to the one expected from backward Brillouin scattering of the plane light wave by a plane acoustic wave. The frequency $\Omega_B$ of the TDBS signal wave packet in Eq.~\eqref{eq:eq15} depends on the interaction angle $\alpha$ between the optical and acoustical beams, the widths of the beams and even the length of the CAP.

\begin{figure}
    \centering
    \includegraphics{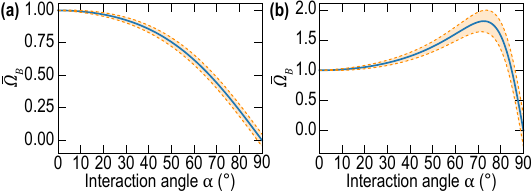}
    \caption{Evolution of the normalized Brillouin (carrier) frequency $\bar{\Omega}_B=\Omega_B/\Omega_B(0)$ and associated normalized spectral full width $\Delta\Omega_B/\Omega_B(0)$ at the $1/e^2$ level as a function of the interaction angle $\alpha$ for (a) $a_\text{probe}/a_\text{\tiny CAP}=1.7$, $l_\text{\tiny CAP}=0$ and (b) $a_\text{probe}/a_\text{\tiny CAP}=0.3$, $l_\text{\tiny CAP}=0$. The full width of the spectral line is related to the wave packet duration via $\Delta\Omega_B=8/\tau_B$ and $\frac{\Delta\Omega_B}{\Omega_B(0)}=\frac{2}{\pi}\frac{\lambda_1}{a_\text{\tiny CAP}}\frac{\sin\alpha}{\sqrt{\left(\frac{a_\text{probe}}{a_\text{\tiny CAP}}\right)^2+\cos^2\alpha+\left(\frac{l_\text{\tiny CAP}}{a_\text{\tiny CAP}}\right)^2\sin^2\alpha}}$, where $\lambda_1$ is the probe wavelength in material (1). In (a) and (b), $\frac{\Delta\Omega_B}{\Omega_B(0)}$ is shown for $\lambda_1/a_\text{\tiny CAP}=0.25$ (and $l_\text{\tiny CAP}=0$).}
    \label{fig:fig3_normalized_BF_vs_alpha}
\end{figure}

\begin{figure}
    \centering
    \includegraphics{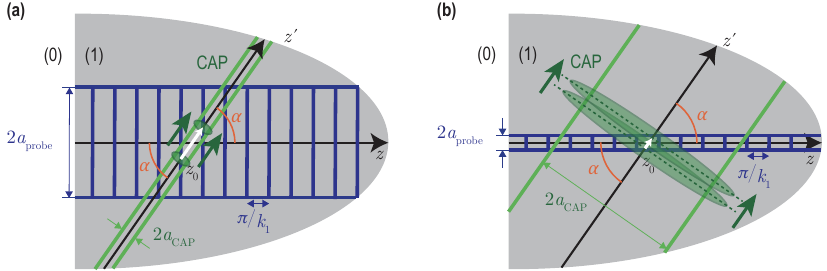}
    \caption{Schemes illustrating the different temporal periodicity of the TDBS processes for two asymptotic relations between the probe light beam radius $a_\text{probe}$ and the CAP beam radius $a_\text{\tiny CAP}$: (a) $a_\text{\tiny CAP}\ll a_\text{probe}$, (b) $a_\text{\tiny CAP}\gg a_\text{probe}$. The propagating CAPs are represented by green ellipses.}
    \label{fig:fig4_temporalperiodicity}
\end{figure}

The analysis demonstrates that, under the condition $a_\text{\tiny CAP}^2<a_\text{probe}^2+2l_\text{\tiny CAP}^2$, the carrier frequency is continuously decreasing with the increase of the interaction angle from $\Omega_B(0)$ at $\alpha=0$ down to 0 for $\alpha=\pi/2$, as illustrated in Fig.~\ref{fig:fig3_normalized_BF_vs_alpha}(a).
However, when the condition $a_\text{\tiny CAP}^2>a_\text{probe}^2+2l_\text{\tiny CAP}^2$ holds, then the carrier frequency, before diminishing down to zero, first grows with increasing $\alpha$ up to its maximal value \[\Omega_B^\text{max}=\frac{1}{2}\left(\sqrt{\frac{a_\text{probe}^2+l_\text{\tiny CAP}^2}{a_\text{\tiny CAP}^2-l_\text{\tiny CAP}^2}}+\sqrt{\frac{a_\text{\tiny CAP}^2-l_\text{\tiny CAP}^2}{a_\text{probe}^2+l_\text{\tiny CAP}^2}}\right)\Omega_B(0)\ge\Omega_B (0)\] at \[\alpha_\text{max}=\sin^{-1}\sqrt{1-\frac{a_\text{probe}^2+l_\text{\tiny CAP}^2}{a_\text{\tiny CAP}^2-l_\text{\tiny CAP}^2}}\le\frac{\pi}{2}\] [Fig.~\ref{fig:fig3_normalized_BF_vs_alpha}(b)]. Particularly, in the case of the CAPs satisfying $l_\text{\tiny CAP}^2\ll a_\text{probe}^2\ll a_\text{\tiny CAP}^2$, which is a rather common condition in picosecond acoustics, the carrier frequency can be much higher than the Brillouin frequency $\Omega_B(0)$: \[\Omega_B^\text{max}\cong \frac{1}{2}\left(\frac{a_\text{probe}}{a_\text{\tiny CAP}} + \frac{a_\text{\tiny CAP}}{a_\text{probe}}\right)\Omega_B(0)\ge\frac{1}{2}\frac{a_\text{\tiny CAP}}{a_\text{probe}}\Omega_B(0)\gg\Omega_B(0).\] The physical sense of the different characteristic frequencies contributing to $\Omega_B$ can be revealed through the analysis of the different limiting cases. It is instructive to analyze the situation that is the closest to most of the experiments reported until nowadays, \textit{i.e.}, $l_\text{\tiny CAP}^2\ll a_\text{probe}^2, a_\text{\tiny CAP}^2$. In this configuration, the biasing influence of the CAP frequency spectrum on the carrier frequency is negligible:
\begin{align}\label{eq:eq17}
    \Omega(l_\text{\tiny CAP}=0)\equiv \Omega_{B0} = \Omega_B(0)\frac{\left(a_\text{probe}^2+a_\text{\tiny CAP}^2\right)\cos\alpha}{a_\text{probe}^2+a_\text{\tiny CAP}^2\cos^2\alpha} 
    \cong \begin{cases}
    \Omega_B(0)\cos\alpha\,,\text{ if }a_\text{\tiny CAP}\ll a_\text{probe}\,,\\
    \frac{\Omega_B(0)}{\cos\alpha}\,,\text{ if }a_\text{\tiny CAP}\gg a_\text{probe}\,.
    \end{cases}
\end{align}
Both characteristic/asymptotic frequencies in Eq.~\eqref{eq:eq17} can be understood by evaluating the temporal periodicity of the acousto-optic interaction process in the considered limiting cases as illustrated in Fig.~\ref{fig:fig4_temporalperiodicity}.

In Fig.~\ref{fig:fig4_temporalperiodicity}, the vertical gratings represented in blue vertical solid lines introduce schematically the sensitivity function \cite{thomsen_coherent_1984, thomsen_surface_1986, grahn_picosecond_1989} $\sin(2k_1z)$ of the probe light to the coherent acoustic field. The spacing between the lines depicts the periodicity of the sensitivity function along the $z$-axis that is $\pi/k_1=\lambda_1/2$, with $\lambda_1$ the optical wavelength of the probe laser beam in medium (1). In Fig.~\ref{fig:fig4_temporalperiodicity}(a), it is clear that the period of the interaction process is equal to the time of flight of the CAP propagating between two blue lines inside the probe light beam. The distance between the blue grating lines along the CAP propagation direction [white solid line arrow along the $z^\prime$-axis in Fig.~\ref{fig:fig4_temporalperiodicity}(a)] is longer than the period $\pi/k_1=\lambda_1/2$, and is equal to $\pi/(k_1\cos\alpha)$. Thus, the period of the process is $T(a_\text{\tiny CAP}\ll a_\text{probe})=\pi/(k_1v_1\cos\alpha)$, while the characteristic frequency is $\Omega_{B0}(a_\text{\tiny CAP}\ll a_\text{probe})=2k_1v_1\cos\alpha=\Omega_B(0)\cos\alpha$. However, in the opposite limiting case $a_\text{\tiny CAP}\gg a_\text{probe}$, the shortest distance between the grating lines along the CAP propagation direction is less than the grating period $\pi/k_1=\lambda_1/2$. It is equal to $(\pi\cos\alpha)/k_1$ [white solid line arrow in Fig.~\ref{fig:fig4_temporalperiodicity}(b)]. Thus, the period of the process is $T(a_\text{\tiny CAP}\gg a_\text{probe})=\pi\cos\alpha/(k_1v_1)$, while the characteristic frequency is $\Omega_{B0}(a_\text{\tiny CAP}\gg a_\text{probe})=2k_1v_1/\cos\alpha=\Omega_B(0)/\cos\alpha$. For all nonzero interaction angles, this characteristic frequency is higher than the so-called \enquote{maximal} frequency of the $180\degree$ backward Brillouin scattering $\Omega_B(0)$. The physical reasons for the difference in the carrier frequencies predicted in the two limiting situations can be additionally appreciated using slightly different wording. In the Fig.~\ref{fig:fig4_temporalperiodicity}(a) case, the different transverse parts of the probe light beam are reflected in the backward direction by the complete CAP beam, which has a phase velocity $v_1\cos\alpha$ along the direction of the probe light beam. In the Fig.~\ref{fig:fig4_temporalperiodicity}(b) case, the complete probe light beam is reflected in the backward direction by the different transverse parts of the CAP beam, which are constituting an effective acoustic mirror moving at the phase velocity $v_1/\cos\alpha$ along the direction of the probe light beam. Thus, the asymptotic frequencies revealed in Eq.~\eqref{eq:eq17} can be interpreted as resulting from the Doppler effect, \textit{i.e.}, from the probe light frequency shift occurring when the probe light is reflected from the moving acoustic \enquote{mirror} \cite{gill_doppler_1965}.

In the general case of Gaussian probe light and CAP beams of arbitrary radii, these two asymptotic characteristic frequencies are contributing to the carrier frequency of the TDBS wave packet via the following formal rule: \[\frac{a_\text{probe}^2+a_\text{\tiny CAP}^2}{\Omega_{B0}}=\frac{a_\text{probe}^2}{\Omega_B(a_\text{\tiny CAP}\ll a_\text{probe})} + \frac{a_\text{\tiny CAP}^2}{\Omega_B(a_\text{\tiny CAP}\gg a_\text{probe})}.\]

\begin{figure}
    \centering
    \includegraphics{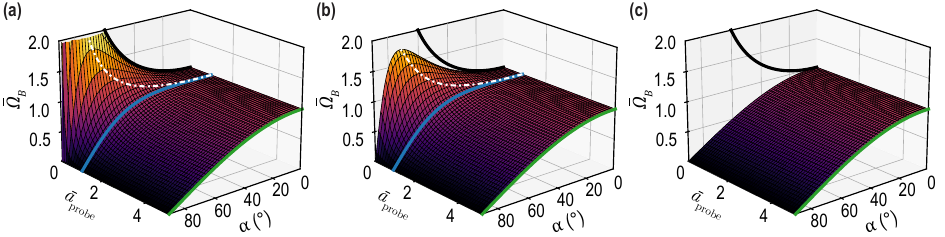}
    \caption{(a-c) Normalized Brillouin (carrier) frequency $\bar{\Omega}_B$ of the wave packet as a function of $\bar{a}_\text{probe}$ and the interaction angle $\alpha$, for three values of $\bar{l}_\text{\tiny CAP}$: (a) 0, (b) 0.3, and (c) 1. In each plot, the asymptotic frequency curves in Eq.~\eqref{eq:eq16} are represented: in black for the case $a_\text{\tiny CAP}\gg a_\text{probe}$ and in green for $a_\text{\tiny CAP}\ll a_\text{probe}$. The blue lines indicate the splitting lines (geometric separators), obtained for $\bar{a}_\text{probe}^2 + 2\bar{l}_\text{\tiny CAP}^2=1$, indicating the boundary between two regimes in (a) and (b). The white dash-dotted lines indicate the maximum position of $\bar{\Omega}_B$ in (a) and (b) where it exists.}
    \label{fig:fig5_3D_BFvariation}
\end{figure}

In Fig.~\ref{fig:fig5_3D_BFvariation}, the normalized carrier frequency $\bar{\Omega}_B(\alpha)\equiv\Omega_B(\alpha)/\Omega_B(0)=\frac{\left(1+\bar{a}_\text{probe}^2\right)\cos\alpha}{\cos^2\alpha+\bar{l}_\text{\tiny CAP}^2\sin^2\alpha+\bar{a}_\text{probe}^2}$ of the TDBS signal [Eq.~\eqref{eq:eq15}] is represented as a function of the interaction angle $\alpha$ and the light and sound beam radii ratio $\bar{a}_\text{probe}=a_\text{probe}/a_\text{\tiny CAP}$ between the probe light and CAP beams, for three different CAP half-length values $\bar{l}_\text{\tiny CAP}=l_\text{\tiny CAP}/a_\text{\tiny CAP}$ (normalized by the CAP radius).

The most interesting result comes from the evolution of the carrier frequency $\Omega_B$: two regimes are evidenced, split by a line (geometric separator) obtained for $\bar{a}_\text{probe}^2 + 2\bar{l}_\text{\tiny CAP}^2=1$ [blue line in Fig.~\ref{fig:fig5_3D_BFvariation}(a-b)]. For $\bar{a}_\text{probe}^2 + 2\bar{l}_\text{\tiny CAP}^2>1$, the normalized Brillouin frequency is only decreasing [Fig.~\ref{fig:fig3_normalized_BF_vs_alpha}(a), for example], while for $\bar{a}_\text{probe}^2 + 2\bar{l}_\text{\tiny CAP}^2<1$, the normalized Brillouin frequency $\bar{\Omega}_B(\alpha)$ increases compared to the value obtained at $\alpha=0\degree$ (here $\bar{\Omega}_B(\alpha=0)=1$) before decreasing to 0 for $\alpha=\pi/2$ [Fig.~\ref{fig:fig3_normalized_BF_vs_alpha}(b), for example]. In Fig.~\ref{fig:fig4_temporalperiodicity}(c), $\bar{l}_\text{\tiny CAP}=1$ and the asymptotic case when $a_\text{\tiny CAP}\ll a_\text{probe}$, $\bar{\Omega}_B(\alpha)=\cos\alpha$, common in picosecond acoustic experiments, is obtained regardless of the value of $\bar{a}_\text{probe}$, \textit{i.e.}, regardless of the ratio of the light and sound beams radii.

The amplitudes of the TDBS wave packets are controlled by the amplitude factor: in the square brackets of Eq.~\eqref{eq:eq15}. It is proportional to the characteristic volume of the CAP, $a_\text{\tiny CAP}^2l_\text{\tiny CAP}$, and the strain pulse magnitude, $\eta_0$. Normalized to magnitude 1 at zero angle, the angular dependent part of this factor
\[
D_B\equiv\frac{\sqrt{a_\text{probe}^2+a_\text{\tiny CAP}^2}}{\sqrt{a_\text{probe}^2+a_\text{\tiny CAP}^2\cos^2\alpha+l_\text{\tiny CAP}^2\sin^2\alpha}}\exp\left[-\frac{k_1^2}{2}\frac{\left(a_\text{\tiny CAP}^2l_\text{\tiny CAP}^2+a_\text{probe}^2l_\text{\tiny CAP}^2\cos^2\alpha+a_\text{probe}^2a_\text{\tiny CAP}^2\sin^2\alpha\right)}{\left(a_\text{probe}^2+a_\text{\tiny CAP}^2\cos^2\alpha+l_\text{\tiny CAP}^2\sin^2\alpha\right)}\right]\,,
\]
can be called the directivity pattern of the TDBS detection (in the specific circumstances of Gaussian probe light and CAP beams). The directivity $D_B$ has an extremum at $\alpha=0$, minimum for $l_\text{\tiny CAP}>a_\text{\tiny CAP}$ and maximum for $l_\text{\tiny CAP}<a_\text{\tiny CAP}$. Thus, the finite length of the CAP could influence the overall shape of the directivity pattern, by diminishing the pre-exponential factor and increasing the exponential factor at non-zero interaction angles. The influence of the $l_\text{\tiny CAP}$ length on the exponential factor contributes to broaden the directivity pattern in comparison with the case of the infinitely short CAPs. In the typical experimental conditions of picosecond ultrasonics ($l_\text{\tiny CAP}\ll a_\text{probe}, a_\text{\tiny CAP}$), the directivity pattern takes a more compact form:
\begin{align}\label{eq:eq_DB0}
    D_B\left(l_\text{\tiny CAP}=0\right)\equiv D_{B0}=\frac{\sqrt{a_\text{probe}^2+a_\text{\tiny CAP}^2}}{\sqrt{a_\text{probe}^2+a_\text{\tiny CAP}^2\cos^2\alpha}}\exp\left[-\frac{k_1^2}{2}\frac{\left(a_\text{probe}^2a_\text{\tiny CAP}^2\sin^2\alpha\right)}{\left(a_\text{probe}^2+a_\text{\tiny CAP}^2\cos^2\alpha\right)}\right]\,.
\end{align}
It is worth mentioning that in comparison with the frequency $\Omega_{B0}$ [Eq.~\eqref{eq:eq17}] the amplitude factor $D_{B0}$ [Eq.~\eqref{eq:eq_DB0}] is always maximum at $\alpha=0$, independently of the ratio of the light and sound beams radii $\bar{a}_\text{probe}$. The strongest dependence on the interaction angle $\alpha$ in the typical experimental conditions of picosecond ultrasonics ($l_\text{\tiny CAP}\ll a_\text{probe}, a_\text{\tiny CAP}$) comes from the Gaussian factor:
\begin{align}\label{eq:eq19}
    \exp\left[-\frac{k_1^2}{2}\frac{\left(a_\text{probe}^2a_\text{\tiny CAP}^2\sin^2\alpha\right)}{\left(a_\text{probe}^2+a_\text{\tiny CAP}^2\cos^2\alpha\right)}\right] = \exp\left[-\frac{k_1^2}{2}\left(\frac{a_\text{probe}^2a_\text{\tiny CAP}^2}{l_B^2}\right)\right] = \exp\left[-2\left(\frac{\pi a_\text{probe}a_\text{\tiny CAP}}{\lambda_1l_B}\right)^2\right]\,.
\end{align}

In TDBS experiments where the additional condition $a_\text{probe}\ll a_\text{\tiny CAP}$ holds, the Gaussian amplitude factor takes the form $\exp\left(-\frac{k_1^2}{2}a_\text{probe}^2\tan^2\alpha\right)$ as far as $\alpha$ is not approaching $\pi/2$. In 3D TDBS imaging experiments, the radii of the pump and probe laser beams are of the same order, \textit{i.e.}, $a_\text{probe}\sim a_\text{\tiny CAP}$ \cite{nikitin_revealing_2015,sandeep_3d_2021,threard_photoacoustic_2021} while both pump and probe pulses are rather strongly focused to increase lateral spatial resolution. The exponential factor in Eq.~\eqref{eq:eq19} is then reduced to $\exp\left(-\frac{k_1^2}{2}\frac{a_\text{\tiny CAP}^2\sin^2\alpha}{1+\cos^2\alpha}\right)$. When $a_\text{probe}\ll a_\text{\tiny CAP}$, the critical inclination angle for which the exponential factor predicts a $1/e^2$ fall in the signal amplitude can be found, from $\frac{k_1^2}{4}a_\text{probe}^2\tan^2\alpha=\left(\frac{\pi a_\text{probe}}{\lambda_1}\tan\alpha\right)^2=1$, in the following form: $\alpha_{1/e^2}=\tan^{-1}\left(\frac{\lambda_1}{\pi a_\text{probe}}\right)$. Assuming that $\lambda_1\le a_\text{probe}$, for a typical material with optical refractive index $n_1\ge 1.5$, the estimate predicts a rather small critical angle: $\alpha_{1/e^2}\cong\frac{1}{\pi n_1}$. When $a_\text{probe}\sim a_\text{\tiny CAP}$, the critical angle could be $\sqrt{2}$ times larger. The analysis of the third case where $a_\text{probe}\gg a_\text{\tiny CAP}$  gives a Gaussian factor of the form $\exp\left(-\frac{k_1^2}{2}a_\text{\tiny CAP}^2\sin^2\alpha\right)$ and thus $\alpha_{1/e^2}=\sin^{-1}\left(\frac{\lambda_1}{\pi a_\text{\tiny CAP}}\right)$. In this case, the angular peak could be potentially broadened if the condition $\lambda_1> a_\text{\tiny CAP}$ is achieved by using, for the optoacoustic generation of CAP, a strongly focused pump laser radiation with a wavelength shorter than the probe light wavelength.

\begin{figure}
    \centering
    \includegraphics{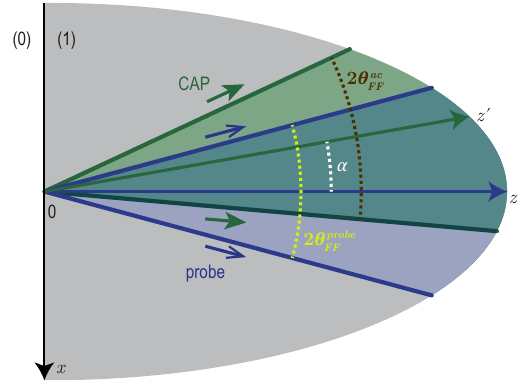}
    \caption{2-D illustration of the probe light and CAP ray distributions (represented as bounded inside cones), highlighting the existence of a critical angle resulting in a strong drop (disappearance) of the TDBS signal amplitude when the collimated light and sound beams are propagating non-collinearly, tilted with an angle $\alpha$. $\theta_{FF}^\text{probe}$ denotes the half angle of divergence of the probe light in the far field, while $\theta_{FF}^\text{ac}$ denotes  the half angle of divergence of the acoustic beam in the far field. At a critical angle $\alpha_{1/e^2}$, the CAP and probe light ray cones stop to overlap.}
    \label{fig:fig6_ray_cones}
\end{figure}

The abrupt diminishing of the scattering efficiency with increasing interaction angle is due to the abrupt diminishing, in the probe light and CAP beams, of the number of quasi-collinearly propagating plane optical and acoustical waves that could satisfy the momentum conservation law in the $180\degree$ backward Brillouin scattering. This conservation law is a prerequisite of an efficient acousto-optic interaction in the case of an acoustically-scattered light heterodyning detection in the opposite direction compared to the incident probe light direction. The existence of a critical angle $\alpha_{1/e^2}$ for the acousto-optic interaction efficiency can be straightforwardly understood from the qualitative analysis of the probe light and CAP beams scattering in the wave vectors space, \textit{i.e.}, via the decomposition of the beams into rays (plane waves). A 2-D representation of the rays’ cones associated to the Gaussian probe light and CAP beams \cite{self_focusing_1983} intersecting/interacting at an angle $\alpha$ is schematically drawn in Fig.~\ref{fig:fig6_ray_cones}.

The probe half angle of divergence obtained in the far field \cite{self_focusing_1983}, $\theta_{FF}^\text{probe}\equiv\frac{\lambda_1}{\pi a_\text{probe}}$, is the critical angle revealed previously in the situation $a_\text{probe}\ll a_\text{\tiny CAP}$. On the other hand, $\theta_{FF}^\text{ac}\equiv\frac{\lambda_\text{ac}}{\pi a_\text{\tiny CAP}}$ is the far field half angle of divergence of the coherent monochromatic acoustic beam with the acoustical wavelength $\lambda_\text{ac}$ and the beam radius $a_\text{CAP}$. The scheme in Fig.~\ref{fig:fig6_ray_cones} illustrates that, as far as there is an overlap between the probe light ray cone and the acoustic ray cone, there will always be plane wave components of the probe light beam to be efficiently scattered by plane wave components of the acoustic beam, in the quasi-collinear $180\degree$ backward Brillouin interaction configuration, corresponding to $\lambda_\text{ac}\cong\frac{\lambda_1}{2}$. The overlap between the cones disappears at the interaction angle $\alpha$ satisfying the geometrical condition $\alpha - \theta_{FF}^\text{ac}=\theta_{FF}^\text{probe}$ (see Fig.~\ref{fig:fig6_ray_cones}). This last relation takes the form $\alpha=\theta_{FF}^\text{probe}+\theta_{FF}^\text{ac}=\frac{\lambda_1}{\pi a_\text{probe}}+\frac{\lambda_1}{2\pi a_\text{\tiny CAP}}$, under the $180\degree$ backscattering condition. It provides qualitatively correct estimates of the critical angles, in Gaussian beams, for all three earlier examined asymptotic situations. Thus, the analysis based on Fig.~\ref{fig:fig6_ray_cones} confirms that the angular directivity in the interaction of probe light and CAP beams is due to the directional selectivity in the efficiency of the TDBS.

From this analysis, it can be expected that, if the CAP is incident on an inclined interface and is reflected/refracted, then the amplitude of the TDBS wave packets will be importantly modified because of possible modifications of the CAP propagation path relatively to that of the probe light. This expectation is further analyzed in Sec.~\ref{subsec:reflection} for the case of the CAP reflection and in Sec.~\ref{subsec:transmission} for that of the CAP transmission/refraction at an inclined material interface.

In summary, the theoretical analysis in Section~\ref{subsec:free_space} predicts that TDBS signals resulting from the interaction of the non-diffracting probe light and CAP Gaussian beams at an angle and far from the material boundaries/interfaces have the form of wave packets. All the parameters of the TDBS wave packets, \textit{i.e.}, the duration, the carrier frequency and the amplitude, depend in general on the radii of the probe light and CAP beams and the duration/length of the CAP.

\subsection{TDBS monitoring of the CAP beam reflection at an interface inclined relatively to the probe light propagation direction}
\label{subsec:reflection}
\begin{figure}
    \centering
    \includegraphics{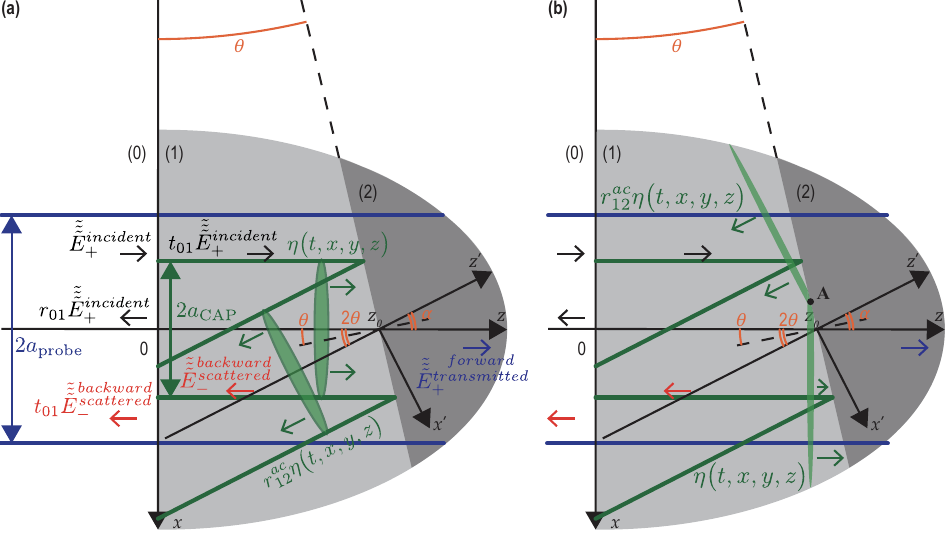}
    \caption{(a) Simplified and (b) detailed sketches of TDBS monitoring of the CAP reflection at the interface between two materials inclined at an angle $\theta$ relatively to the free surface ($z=0$) of material (1) ($z>0$). The deviation angle, with which the central ray of the coherent acoustic beam deviates from its initial propagation direction, is equal to $2\theta$. The probe light beam is neither refracted nor reflected by the considered material interface since both materials have the same optical properties. Continuous green and blue lines present, inside their Rayleigh ranges, the acoustic and laser beams, respectively. The CAPs are additionally presented by green ellipses to illustrate their strong localization along the propagation direction caused by their ultrashort duration. In (b), the simplified CAP beam representation of finite $2a_\text{\tiny CAP}$ lateral distribution shown in (a) is made more accurate to account for the infinite lateral dimension of the Gaussian CAP beam.}
    \label{fig:fig7_scheme_reflection}
\end{figure}

In Fig.~\ref{fig:fig7_scheme_reflection}, the reflection of a CAP at the interface between two different materials is presented. The materials are labeled (1) and (2) and are assumed homogeneous and isotropic with the same optical and acousto-optical, but different elastic properties. Thus, contrary to the CAP beam, the probe light beam is neither reflected nor refracted by the assumed material interface. Note that the possible reflected/transmitted shear CAP, that could emerge as a result of the mode conversion of the longitudinal CAP incident to the inclined material interface, is not presented in Fig.~\ref{fig:fig7_scheme_reflection}. Moreover, in the development of the theory in Sec.~\ref{sec:TDBS_arbitratry_angle}, the possible multimode nature of the acoustical and optical fields is persistently neglected. Please refer to Sec.~\ref{sec:discussion} for a discussion on the validity of these assumptions and the comments on the theory modifications perspectives to avoid them.

Figure~\ref{fig:fig7_scheme_reflection}(a) schematically presents the CAP before and after reflection from the material interface of its central part, represented by the cylinders of the characteristic diameter $2a_\text{\tiny CAP}$. However, to evaluate the transient processes, we need to take explicitly into account that the acoustic field of the CAP is distributed through its complete radially infinite Gaussian beam, and is non-zero outside the cross section of the represented cylinders. The reflection of the CAP from the interface is sketched in Fig.~\ref{fig:fig7_scheme_reflection}(b) with more details. Note that the CAP is not shown in Fig.~\ref{fig:fig7_scheme_reflection}(b), contrary to Fig.~\ref{fig:fig7_scheme_reflection}(a), before and after reflection, but at a given instant when a part of the beam has already reflected on the interface while another part has not encountered the interface yet.

Starting from here, we only consider the typical picosecond acoustic situation where, for revealing the essential features of the Brillouin scattering, it is sufficient to model the CAP as a strain that is delta-localized along the CAP propagation direction \cite{akhmanov_laser_1992,gusev_advances_2018,gusev_depth-profiling_2011}. Then, from the geometry of the CAP reflection in Fig.~\ref{fig:fig7_scheme_reflection}(b), it follows that the reflected CAP is localized in the half-plane above the x-coordinate of point A while the incident CAP is localized in the spatial region below the x-coordinate of point A. This point A stands for the current position of the incident CAP at the interface between materials (1) and (2). Note that Fig. 7 is prepared for interfaces inclined with an angle $\theta<\pi/4$ relatively to the free surface of material (1) ($z=0$). For larger inclinations, both the incident CAP and the reflected CAP exist only in the spatial region below the x-coordinate of point A. The strain fields in the incident and reflected CAPs are described by:
\begin{align*}
    \eta_\text{incident}(t,x,y,z)&=\eta_0\sqrt{\frac{\pi}{2}}l_\text{\tiny CAP}e^{-2\frac{x^2+y^2}{a^2_\text{\tiny CAP}}}\delta(z-z_0-v_1t)\,, \\
    \eta_\text{reflected}(t,x^\prime,y^\prime,z^\prime)&=r_{12}^{ac}(v_1, v_2, \theta)\eta_0\sqrt{\frac{\pi}{2}}l_\text{\tiny CAP}e^{-2\frac{x^{\prime 2}+y^{\prime 2}}{a^2_\text{\tiny CAP}}}\delta(z^\prime+v_1t)\,, \\
    &=r_{12}^{ac}(v_1, v_2, \theta)\eta_0\sqrt{\frac{\pi}{2}}l_\text{\tiny CAP}e^{-2\frac{\left[x\cos\alpha+(z-z_0)\sin\alpha\right]^2+y^2}{a^2_\text{\tiny CAP}}}\delta(-x\sin\alpha+(z-z_0)\cos\alpha+v_1t)\,.
\end{align*}
In these equations, $l_\text{\tiny CAP}$ stands for the half characteristic length of the ultrashort CAP and $r_{12}^{ac}(v_1, v_2, \theta)$ denotes the acoustic reflection coefficient at the interface. Note that the acoustic reflection and transmission  coefficients depend not only on the velocities $v_1$ and $v_2$ of the longitudinal (compression -- dilatation) acoustic waves in the materials (1) and (2), but also on their densities and the velocities of the transversal (shear) acoustic waves. Above and in the following, we are not writing these dependencies explicitly for the sake of compactness. The angle between the incident and reflected CAPs propagation directions defined in Fig.~\ref{fig:fig7_scheme_reflection} is $\alpha=2\theta$, while the zero instant of time is associated to the arrival of the delta-CAP central ray at $z=z_0$. The $x$ coordinate of point A is thus described by $x_A(t,v_1,\theta)=\frac{v_1 t}{\tan\theta}$. The application of Eq.~\eqref{eq:eq14}, with the integration over the $x$ coordinate limited to $x_A(t,v_1,\theta)\le x\le +\infty$, leads to the description of the contribution to the TDBS signal from the incident CAP:
\begin{align}\label{eq:eq20}
    \left(\frac{\mathrm{d}R}{R}\right)_\text{incident}\equiv B_i = P\sqrt{\frac{\pi}{2}}\frac{\eta_0l_\text{\tiny CAP}a_\text{\tiny CAP}^2}{a_\text{probe}^2+a_\text{\tiny CAP}^2}\Biggl\langle\frac{1}{2}\left[1-\mathrm{erf}\left(\sqrt{2}\frac{x_A(t)}{a_\text{overlap}}\right)\right]\Biggr\rangle\sin\left[\Omega_B(0)\left(t+\frac{z_0}{v_1}\right)\right]\,,
\end{align}
where a compact notation $a_\text{overlap}$ is introduced for the characteristic spatial scale defined by the following \enquote{summation} rule: $\frac{1}{a_\text{overlap}^2}\equiv \frac{1}{a_\text{probe}^2} + \frac{1}{a_\text{\tiny CAP}^2}$. The incident CAP is collinear to the probe light, \textit{i.e.}, the interaction angle $\alpha=0$. Thus, the Gaussian amplitude factor in Eq.~\eqref{eq:eq19}, derived earlier for the interaction of beams in homogeneous isotropic material, is equal to 1, while the carrier frequency in Eq.~\eqref{eq:eq16} is reduced to $\Omega_B(0)$ in Eq.~\eqref{eq:eq20}. In the incident CAP contribution to the TDBS signal $B_i$ [Eq.~\eqref{eq:eq20}], the wave packet envelope of Eq.~\eqref{eq:eq15}, associated with the finite length $l_B$ [Eq.~\eqref{eq:eq16}] of the intersection volume of light and sound beams, is absent. This is because the characteristic length $l_B$ diverges for $\alpha=0$, \textit{i.e.}, $l_B(\alpha=0)=+\infty$ [see Fig.~\ref{fig:fig4_temporalperiodicity}(a-b)]. The factor $\left[1-\mathrm{erf}\left(\sqrt{2}\frac{x_A(t)}{a_\text{overlap}}\right)\right]$ in Eq.~\eqref{eq:eq20} describes the time decay of the incident CAP contribution to the signal when the incident pulse is \enquote{disappearing} via its transformation into reflected and transmitted CAPs:
\begin{align}\label{eq:eq21}
    \left[1-\mathrm{erf}\left(\sqrt{2}\frac{x_A(t)}{a_\text{overlap}}\right)\right] = \left[1-\mathrm{erf}\left(\sqrt{2}\frac{v_1t}{a_\text{overlap}\tan\theta}\right)\right]\equiv\left[1-\mathrm{erf}\left(\sqrt{2}\frac{t}{\tau_i}\right)\right]\,.
\end{align}

As the above definition of $a_\text{overlap}$ indicates, it corresponds to the radius at $1/e^2$ level of the overlap of the probe light intensity and CAP amplitude Gaussian distributions. The term $a_\text{overlap}\tan\theta$ is therefore clearly defining the characteristic distance along the CAP propagation direction of the cylindrical volume where the delta CAP, the probe light and the interface could coexist when the CAP is being reflected by the interface. In other words, this is an estimate of the distance that the CAP should travel between the characteristic moments of reaching the interface for the first time and losing the contact with it, as it is monitored by TDBS of the probe light beam. The physical sense of the characteristic time $\tau_i=\frac{a_\text{overlap}\tan\theta}{v_1}$ in Eq.~\eqref{eq:eq21} can be understood as the time, monitored via TDBS, taken by the incident CAP to be reflected/transmitted at the interface. Note that this time is always longer than the reflection/transmission time that is defined only by the radius of the CAP, \textit{i.e.}, omitting the role of the probe light beam radius in the TDBS detection.

\begin{figure}
    \centering
    \includegraphics{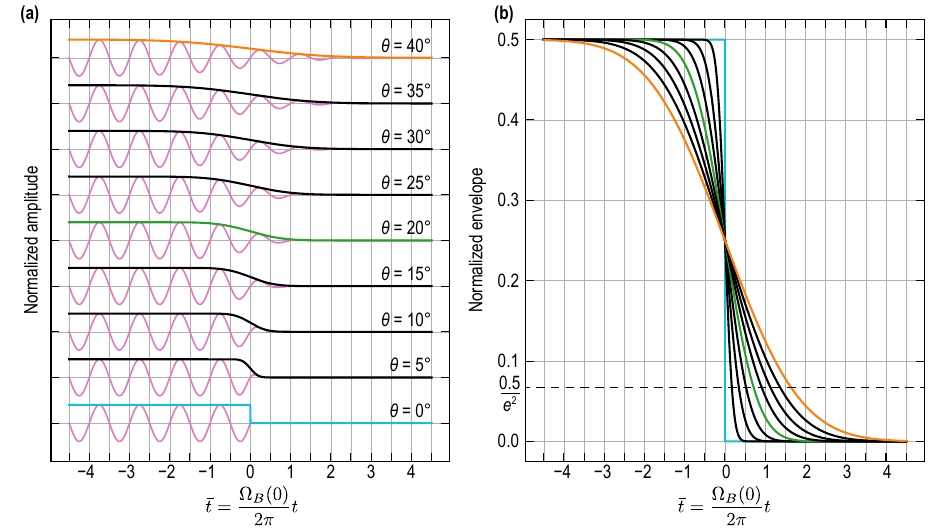}
    \caption{Contribution to the TDBS signal from the incident CAP [Eq.~\eqref{eq:eq20}] as a function of the interface inclination angle $\theta$.
    (a) Normalized $\left(\frac{\mathrm{d}R}{R}\right)_\text{incident}$ (solid oscillating pink curves) for $\theta\in[0\si{\degree},40\si{\degree}]$ with 5\si{\degree} steps, and associated envelopes (in solid black lines + 3 colored lines). The contributions are upshifted for better visualization. (b) Normalized $\left(\frac{\mathrm{d}R}{R}\right)_\text{incident}$ envelopes (in solid black lines + 3 colored lines) for $\theta\in[0\si{\degree},40\si{\degree}]$ with 5\si{\degree} steps.}
    \label{fig:fig8_incCAP_contrib}
\end{figure}

At this point, the discussion is sufficiently developed to introduce 
an illustrative example. Figure~\ref{fig:fig8_incCAP_contrib}(a) provides the contributions of the incident CAP to the TDBS signal [Eq.~\eqref{eq:eq20}] (solid oscillating pink curves) along with their amplitude envelopes (colored and black curves), as a function of the inclination $\theta$ of the interface. The time axis is normalized with the Brillouin frequency at zero interaction angle in material (1): $\bar{t}=\Omega_B(0)/(2\pi)t=2v_1/\lambda_1 t$. The plotted signals and envelopes are normalized by $P\sqrt{\frac{\pi}{2}}\eta_0l_\text{\tiny CAP}$. Materials (1) and (2) are assumed to be of cubic symmetry which 
is relevant since it is optically isotropic (not birefringent) but elastically anisotropic. This case can either depict an inclined interface between two cubic materials or an inclined intergrain interface in a cubic polycrystalline material
as illustrated in Fig.~\ref{fig:fig7_scheme_reflection}. 
Only the propagation of a longitudinal CAP (LA mode) is here considered for the computation of Eq.~\eqref{eq:eq20}. In Fig.~\ref{fig:fig8_incCAP_contrib}(b) the envelopes of the $\frac{\mathrm{d}R}{R}$ wave packets related to the incident CAP are presented separately for better appreciation. Three specific envelopes are highlighted with colors, instead of using black lines in Fig.~\ref{fig:fig8_incCAP_contrib}, corresponding to the cases $\theta=0$\si{\degree} (blue), $\theta=20$\si{\degree} (green) and $\theta=40$\si{\degree} (orange). By normalizing all spatial variables in Eq.~\eqref{eq:eq20} by the probe beam radius $a_\text{probe}$ and the normalized time $\bar{t}$, it can be easily shown that the expression of $\left(\frac{\mathrm{d}R}{R}\right)_\text{incident}$ depends on three nondimensionalized parameters: $a_\text{\tiny CAP}/a_\text{probe}$, $2a_\text{probe}/\lambda_1$, and $2z_0/\lambda_1$. In Fig.~\ref{fig:fig8_incCAP_contrib}, the first two parameters were set to values relevant to experimental conditions, $a_\text{\tiny CAP}/a_\text{probe}=1$ and $2a_\text{probe}/\lambda_1=5$, while $2z_0/\lambda_1$ was kept at zero.

In the case of an interface parallel to the free surface of material~(1), \textit{i.e.}, parallel to the $z=0$ plane and with $\theta=0$\si{\degree}, the contribution of the incident CAP to the full TDBS signal is stopped/ended when the CAP reaches the interface (blue envelope in Fig.~\ref{fig:fig8_incCAP_contrib}), as it is reflected and transmitted at the interface. We recall that the zero instant of time, $\bar{t}=0$, is associated to the arrival of the delta-CAP central ray at $z=z_0$. For small interface inclination angles $\theta$, the contribution of the incident CAP to the TDBS signal decreases abruptly. For example, when the boundary is inclined with an angle $\theta=20$\si{\degree}, the amplitude of the envelope reaches the $1/e^2$ level before $\bar{t}=1$, that is to say in less than a period of Brillouin oscillation (green envelope in Fig.~\ref{fig:fig8_incCAP_contrib}).
This contribution broadens with increasing inclination of the interface as expressed by the process of reflection/transmission of the pulse in Fig.~\ref{fig:fig7_scheme_reflection}(b): the progressive decrease of the envelope starts before the arrival of the delta-CAP central ray at $z=z_0$ (before $\bar{t}=0$), but reaches the $1/e^2$ level few Brillouin periods later. Note that changing the value of $2z_0/\lambda_1$ only results in changing the phase of the TDBS signal and that any integer multiple of $z_0/\lambda_1$ leads to zero amplitude of the TDBS signal at $\bar{t}=0$.

The contribution to the TDBS signal from the reflected pulse is a bit more cumbersome and is obtained from Eq.~\eqref{eq:eq14} with the integration over the $x$ coordinate limited this time to $-\infty\le x\le x_A(t,v_1,\theta)$:
\begin{align}\label{eq:eq22}
\begin{split}
    &\left(\frac{\mathrm{d}R}{R}\right)_\text{refl.} \equiv B_r = r_{12}^{ac}\sqrt{\frac{\pi}{2}}P{\footnotesize\frac{\eta_0l_\text{\tiny CAP}a_\text{\tiny CAP}^2\cos\alpha}{\sqrt{a_\text{probe}^2+a_\text{\tiny CAP}^2}\sqrt{a_\text{probe}^2+a_\text{\tiny CAP}^2\cos^2\alpha}}} e^{-\frac{k_1^2}{2}\frac{a_\text{probe}^2a_\text{\tiny CAP}^2\sin^2\alpha}{a_\text{probe}^2+a_\text{\tiny CAP}^2\cos^2\alpha}} e^{-2\frac{\sin^2\alpha(v_1t)^2}{a_\text{probe}^2+a_\text{\tiny CAP}^2\cos^2\alpha}}\\ 
    \times& \Im\left\{e^{-i\Omega_B(0)\left[\frac{\left(a_\text{probe}^2+a_\text{\tiny CAP}^2\right)\cos\alpha}{a_\text{probe}^2+a_\text{\tiny CAP}^2\cos^2\alpha}t-\frac{z_0}{v_1}\right]}\dots \right. \\ & \left.\qquad\qquad\dots\times\Biggl\langle\frac{1}{2}\left[1+\mathrm{erf}\left({\footnotesize\frac{\sqrt{2}x_A(t)}{a_{\text{overlap},\alpha}}}-{\footnotesize\frac{\sqrt{2}\sin\alpha v_1t}{a_\text{\tiny CAP}^2\cos^2\alpha}}a_{\text{overlap},\alpha}-i\sqrt{\frac{k_1^2}{2}{\footnotesize\frac{a_\text{probe}^2a_\text{\tiny CAP}^2\sin^2\alpha}{a_\text{probe}^2+a_\text{\tiny CAP}^2\cos^2\alpha}}}\right)\right]\Biggr\rangle\right\}\,,
\end{split}
\end{align}
where a compact notation $a_{\text{overlap},\alpha}$, similar to that in Eq.~\eqref{eq:eq20}, is introduced for the characteristic spatial scale defined by the following \enquote{summation} rule: $\frac{1}{a_{\text{overlap},\alpha}^2}\equiv \frac{1}{a_\text{probe}^2} + \frac{1}{a_\text{\tiny CAP}^2\cos^2\alpha}$. The similarity with Eq.~\eqref{eq:eq20} means that the roles of most of the terms in Eq.~\eqref{eq:eq22} have been already discussed. We can immediately recognize the Gaussian envelope of the wave packet, the oscillation (carrier) frequency and the angle-dependent constant in the time-dependent Gaussian amplitude factor derived earlier in Sec.~\ref{subsec:free_space} for the interaction between probe light and CAP beams in isotropic homogeneous materials [Eq.~\eqref{eq:eq15}]. The contribution from the reflected CAP to the TDBS signal [Eq.~\eqref{eq:eq22}] differs from Eq.~\eqref{eq:eq15} obtained in Sec.~\ref{subsec:free_space} via the addition of spatio-temporal scales, regrouped in the factor $G_r$:
\begin{align}\label{eq:eq23}
    G_r\equiv\left[1+\mathrm{erf}\left({\footnotesize\frac{\sqrt{2}x_A(t)}{a_{\text{overlap},\alpha}}}-{\footnotesize\frac{\sqrt{2}\sin\alpha v_1t}{a_\text{\tiny CAP}^2\cos^2\alpha}}a_{\text{overlap},\alpha}-i\sqrt{\frac{k_1^2}{2}{\footnotesize\frac{a_\text{probe}^2a_\text{\tiny CAP}^2\sin^2\alpha}{a_\text{probe}^2+a_\text{\tiny CAP}^2\cos^2\alpha}}}\right)\right]\,.
\end{align}
This factor contributes to the description of the dynamics of \enquote{generation} of the reflected CAP by the incident one, when monitored by TDBS. The first term in the argument of the error function, denoted $\mathrm{erf}$, contributes to the description of the appearance/disappearance dynamics of the reflected pulse in the characteristic volume of the probe light field near the inclined interface. The definition of $a_{\text{overlap},\alpha}$ indicates that it corresponds to the radius at $1/e^2$ level of the overlap of the probe light intensity and CAP amplitude Gaussian distributions along the $x$-axis. $a_{\text{overlap},\alpha}$ differs from $a_{\text{overlap}}$ only by replacing the incident CAP beam radius with the reflected CAP beam radius, both projected on the $x$-axis. Thus, the sense of the characteristic time $\tau_r=\frac{a_{\text{overlap},\alpha}\tan\theta}{v_1}$ in the first term of the $\mathrm{erf}$ function argument in Eq.~\eqref{eq:eq23} is similar to the already discussed sense of $\tau_i=\frac{a_{\text{overlap}}\tan\theta}{v_1}$. The second time-dependent term in the argument of the $\mathrm{erf}$ function could be associated with the time $\tau_B(l_\text{\tiny CAP}=0)$ of the CAP propagation across the probe light intensity and the reflected CAP overlap region via the following relation:
\[\frac{\sin\alpha v_1t}{a_\text{\tiny CAP}^2\cos^2\alpha}a_{\text{overlap},\alpha}=\frac{a_\text{probe}}{a_\text{\tiny CAP}}\frac{v_1t}{l_B(l_\text{\tiny CAP}=0)}=\frac{a_\text{probe}}{a_\text{\tiny CAP}}\frac{t}{\tau_B(l_\text{\tiny CAP}=0)}\,,\]
where $l_B(l_\text{\tiny CAP}=0)=\sqrt{\left(\frac{a_\text{probe}}{\sin\alpha}\right)^2+\left(\frac{a_\text{\tiny CAP}}{\tan\alpha}\right)^2}$ from Eq.~\eqref{eq:eq16}. The dynamics of the $G_r$ factor described by Eq.~\eqref{eq:eq23} drastically depends on the sign of the combined characteristic time $\bar{\tau}$ in the argument of the $\mathrm{erf}$ function:
\[\frac{\sqrt{2}x_A(t)}{a_{\text{overlap},\alpha}}-\frac{\sqrt{2}\sin\alpha v_1t}{a_\text{\tiny CAP}^2\cos^2\alpha}a_{\text{overlap},\alpha}\equiv \frac{t}{\tau_r}-\frac{a_\text{probe}}{a_\text{\tiny CAP}}\frac{t}{\tau_B(l_\text{\tiny CAP}=0)}=\frac{t}{\bar{\tau}}\,.\]
If $\bar{\tau}>0$, the amplitude of $G_r$ is growing from 0 at $t=-\infty$ to 2 at $t=+\infty$, while, if $\bar{\tau}<0$, then its dynamics is reverse and it decreases from 2 at $t=-\infty$ to 0 at $t=+\infty$.

\begin{figure}
    \centering
    \includegraphics{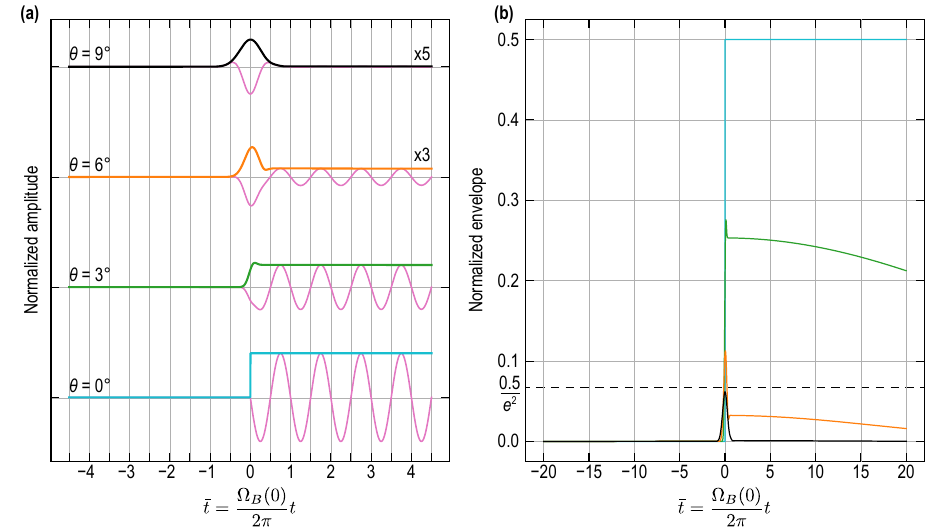}
    \caption{Contribution to the TDBS signal from the reflected CAP [Eq.~\eqref{eq:eq22}] as a function of the interface inclination angle $\theta$.
    (a) Normalized $\left(\frac{\mathrm{d}R}{R}\right)_\text{refl.}$ (solid oscillating pink curves) for $\theta\in\{0\si{\degree}, 3\si{\degree}, 6\si{\degree}, 9\si{\degree}\}$, and associated envelopes. The contributions are upshifted for better visualization. (b) Normalized $\left(\frac{\mathrm{d}R}{R}\right)_\text{refl.}$ envelopes for $\theta\in\{0\si{\degree}, 3\si{\degree}, 6\si{\degree}, 9\si{\degree}\}$.} 
    \label{fig:fig10_old_reflCAP_contrib}
\end{figure}

Figure~\ref{fig:fig10_old_reflCAP_contrib}(a) provides the contribution of the reflected CAP to the TDBS signal (solid oscillating pink curves) along with their amplitude envelopes, as a function of the inclination $\theta$ of the interface.
The same normalization of the time axis and set of nondimensionalized parameters as for the case of the incident CAP were used. 
In Fig.~\ref{fig:fig10_old_reflCAP_contrib}(b) the envelopes of the $\left(\frac{\mathrm{d}R}{R}\right)_\text{refl.}$ wave packets related to the reflected CAP are presented separately with a larger range of values for $\bar{t}$ in order to better appreciate the gradual amplitude decrease for better appreciation. Four specific cases are presented in Fig.~\ref{fig:fig10_old_reflCAP_contrib}, corresponding to the inclination angles $\theta=0$~\si{\degree} (blue), $\theta=3$~\si{\degree} (green), $\theta=6$~\si{\degree} (orange) and $\theta=9$~\si{\degree} (black). In Fig.~\ref{fig:fig10_old_reflCAP_contrib}(a), the amplitude of the TDBS signals and envelopes are multiplied by a factor of 3 and 5 for ease of observation of the case $\theta=6$~\si{\degree} and $\theta=9$~\si{\degree}, respectively. The plotted TDBS signals and envelopes are normalized by $Pr_{12}^{ac}\sqrt{\frac{\pi}{2}}\eta_0l_\text{\tiny CAP}$.

In agreement with Fig.~\ref{fig:fig8_incCAP_contrib}, in the case of an interface parallel to the free surface of material (1), \textit{i.e.}, $\theta=0$\si{\degree}, the contribution of the reflected CAP to the full TDBS signal is starting/emerging when the CAP reaches the interface (blue envelope in Fig.~\ref{fig:fig10_old_reflCAP_contrib}). Then, the contribution of the reflected CAP falls abruptly with increasing inclination angle. For example, for $\theta=6$\si{\degree}, the amplitude of the envelope falls to the $1/e^2$ level quasi-instantly (orange line in Fig.\ref{fig:fig10_old_reflCAP_contrib}). Thus, it is safe to say that the contribution of the reflected CAP to the TDBS signal is limited to the vicinity of the interface, and is negligible when the interface is inclined, even for small angles. The reason is found in the discussion related to Fig.~\ref{fig:fig6_ray_cones}: the deviation angle ($2\theta$) of the propagation direction of the reflected CAP relative to the probe propagation direction (Fig.~\ref{fig:fig7_scheme_reflection}) is such that the number of quasi-collinearly propagating photons and phonons that could satisfy the momentum conservation law in the 180° backward Brillouin scattering is abruptly diminished. In Fig.~\ref{fig:fig10_old_reflCAP_contrib}, it is also evident that a distinct temporal behavior (feature) emerges around $\bar{t}=0$ for non-zero inclination angles. This feature becomes more apparent when examining the envelope, but it can also be observed in the TDBS signals (shown in pink) in Fig.~\ref{fig:fig10_old_reflCAP_contrib}(a). This contribution arises from the reflection of the $\delta$-CAP on the inclined interface in addition to the Brillouin oscillations. While the CAP is assumed to be $\delta$-localized in space along its propagation direction, it possesses a lateral Gaussian distribution that introduces a characteristic time scale associated with the process of CAP reflection. Hence, the characteristic time of this CAP/interface interaction feature depends on the widths of the probe light and acoustic beams, as well as the inclination angle $\theta$. Larger beam widths or a larger $\theta$ result in a longer duration for the the process of the $\delta$-CAP reflection at the interface. Finally, the nondimensionalized parameter $2z_0/\lambda_1$ plays here also a role on the phase of the TDBS signals. Note that, to prepare these illustrations, we neglected the elastic anisotropy of the cubic materials. The discussion on the influence of elastic anisotropy on the reflected CAP is presented in Section~\ref{sec:discussion}.

\subsection{TDBS monitoring of the CAP beam transmission across a material interface inclined relatively to the probe light propagation direction}
\label{subsec:transmission}
\begin{figure}
    \centering
    \includegraphics{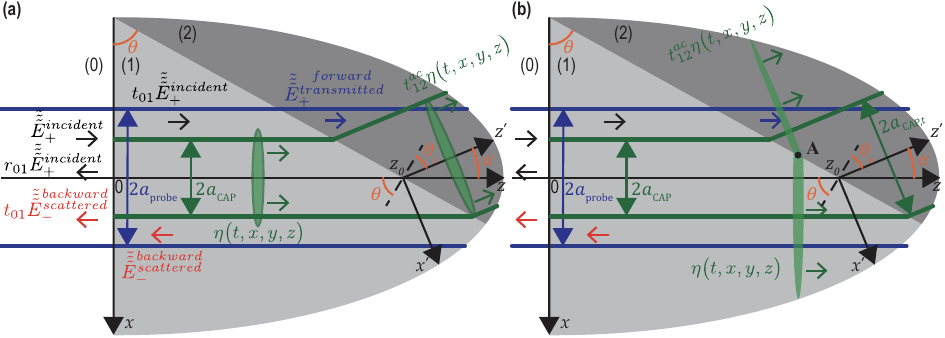}
    \caption{(a) Simplified and (b) detailed sketches of TDBS monitoring of the CAP transmission across the interface between two materials inclined at an angle $\theta$ relative to the free surface ($z = 0$) of material (1) ($z>0$) in the case $v_1>v_2$. Continuous green and blue lines present, inside their Rayleigh ranges, the acoustic and laser beams, respectively. The deviation angle, with which the central ray of the coherent acoustic beam deviates from its initial propagation direction, is denoted by $\alpha$. The diameter of the Gaussian CAP beam changes from $2a_\text{\tiny CAP}$ to $2a_{\text{\tiny CAP}, t}$ in the transmission process. The probe light beam is neither refracted nor reflected by the considered material interface since both materials have the same optical properties. The CAPs are additionally presented by green ellipses to illustrate their strong localization along the propagation direction caused by their ultrashort duration. In (b), the simplified CAP beam representation of finite lateral distribution shown in (a) is made more accurate to account for the infinite lateral dimension of the Gaussian CAP beam, both before and after transmission.}
    \label{fig:fig11_old_scheme_transmission}
\end{figure}

In Fig.~\ref{fig:fig11_old_scheme_transmission}, the transmission of a CAP across the interface between two materials is presented. The two materials are labeled (1) and (2) and are assumed to be homogeneous and isotropic with the same optical and acousto-optical, but different acoustical properties. The other assumptions assumed in Fig.~\ref{fig:fig11_old_scheme_transmission} are the same as those for Fig.~\ref{fig:fig7_scheme_reflection}.

Figure~\ref{fig:fig11_old_scheme_transmission}(a) schematically presents the CAP before and after transmission, across the interface between two materials, of its central part represented by the cylinders of characteristic radii $a_\text{\tiny CAP}$ to $a_{\text{\tiny CAP}, t}$, respectively. However, to evaluate the transient processes, we need to take into account explicitly that the acoustic field of the CAP is distributed in its complete Gaussian beam and is thus non-zero outside the cross section of the presented cylinders. For this purpose, the transmission process is sketched with more details in Fig.~\ref{fig:fig11_old_scheme_transmission}(b). The transmitted acoustic pulse is described by:
\begin{align}\label{eq:eq25}
    \eta_\text{transmitted}&(t,x^\prime,y^\prime,z^\prime)=t_{12}^{ac}(v_1, v_2, \theta)\eta_0\sqrt{\frac{\pi}{2}}l_{\text{\tiny CAP},t}e^{-2\frac{x^{\prime 2}}{a^2_{\text{\tiny CAP},t}}}e^{-2\frac{y^{\prime 2}}{a^2_{\text{\tiny CAP}}}}\delta(z^\prime-v_2t)\,, \nonumber\\
    &=t_{12}^{ac}(v_1, v_2, \theta)\eta_0\sqrt{\frac{\pi}{2}}l_{\text{\tiny CAP},t}e^{-2\frac{\left[x\cos\alpha+(z-z_0)\sin\alpha\right]^2}{a^2_{\text{\tiny CAP},t}}}e^{-2\frac{y^2}{a^2_{\text{\tiny CAP}}}}\delta(-x\sin\alpha+(z-z_0)\cos\alpha-v_2t)\,.
\end{align}
In Eq.~\eqref{eq:eq25}, $v_2$ is the acoustic velocity in the material denoted (2), $2l_{\text{\tiny CAP},t}=2l_{\text{\tiny CAP}}(v_2/v_1)$ stands for the characteristic length of the ultrashort CAP in the second material, $t_{12}^{ac}(v_1,v_2,\theta)$ compactly denotes the acoustic transmission coefficient at the interface between the two materials (1) and (2), while the zero instant of time is associated with the arrival of the delta-CAP central ray at the distance $z=z_0$ from the surface $z=0$. We recall here that the acoustic transmission coefficient depends also on the densities and shear acoustic velocities of both materials. The angle between the incident and transmitted CAPs propagation directions in Fig.~\ref{fig:fig11_old_scheme_transmission} is $\alpha=\theta-\theta_t$, where $\theta$ and $\theta_t$ are commonly defined as the angles of CAP incidence and CAP transmission/refraction, respectively. It is worth mentioning that Eq.~\eqref{eq:eq25} takes into account the modification in the projection of the transmitted CAP cross section on the x-axis: $a_{\text{\tiny CAP},t}=a_\text{\tiny CAP}(\cos\theta_t/\cos\theta)$.

Figure~\ref{fig:fig11_old_scheme_transmission} was drawn under the particular condition $(v_2/v_1)<1$. In Fig.~\ref{fig:fig11_old_scheme_transmission}(b), it is demonstrated that, under this particular condition, the transmitted CAP is located in the spatial region $-\infty\le x\le x_A(t, v_1, \theta)$, where the integration in Eq.~\eqref{eq:eq14} should be done for the evaluation of the transmitted CAP contribution to the TDBS signal. Finally, the contribution to the TDBS signal from the transmitted CAP reads:
\begin{align}\label{eq:eq26}
\begin{split}
    &\left(\frac{\mathrm{d}R}{R}\right)_\text{trans.} \equiv B_t = t_{12}^{ac}\sqrt{\frac{\pi}{2}}P{\footnotesize\frac{\eta_0l_\text{\tiny CAP}a_\text{\tiny CAP}a_{\text{\tiny CAP},t}\cos\alpha}{\sqrt{a_\text{probe}^2+a_\text{\tiny CAP}^2}\sqrt{a_\text{probe}^2+a_{\text{\tiny CAP},t}^2\cos^2\alpha}}} \\ 
    \times & e^{-\frac{k_1^2}{2}\frac{a_\text{probe}^2a_{\text{\tiny CAP},t}^2\sin^2\alpha}{a_\text{probe}^2+a_{\text{\tiny CAP},t}^2\cos^2\alpha}} e^{-2\frac{\sin^2\alpha(v_2t)^2}{a_\text{probe}^2+a_{\text{\tiny CAP},t}^2\cos^2\alpha}} \Im\left\{e^{+i\Omega_B(0)\frac{v_2}{v_1}\left[\frac{\left(a_\text{probe}^2+a_{\text{\tiny CAP},t}^2\right)\cos\alpha}{a_\text{probe}^2+a_{\text{\tiny CAP},t}^2\cos^2\alpha}t+\frac{z_0}{v_2}\right]}\dots \right. \\ & \left.\qquad\quad\dots\times\Biggl\langle\frac{1}{2}\left[1+\mathrm{erf}\left({\footnotesize\frac{\sqrt{2}x_A(t)}{a_{\text{overlap},\alpha,t}}}+{\footnotesize\frac{\sqrt{2}\sin\alpha v_2t}{a_{\text{\tiny CAP},t}^2\cos^2\alpha}}a_{\text{overlap},\alpha,t}-i\sqrt{\frac{k_1^2}{2}{\footnotesize\frac{a_\text{probe}^2a_{\text{\tiny CAP},t}^2\sin^2\alpha}{a_\text{probe}^2+a_{\text{\tiny CAP},t}^2\cos^2\alpha}}}\right)\right]\Biggr\rangle\right\}\,,
\end{split}
\end{align}

The structure of the transmitted CAP contribution $\left(\frac{\mathrm{d}R}{R}\right)_\text{trans.}$ in Eq.~\eqref{eq:eq26}, in the case $(v_2/v_1)<1$ (Fig.~\ref{fig:fig11_old_scheme_transmission}) is the same as the structure of the reflected CAP contribution $\left(\frac{\mathrm{d}R}{R}\right)_\text{refl.}$ in Eq.~\eqref{eq:eq22}. The involved quantities are just scaled from those in material (1) to those in material (2) because of the transmission of the CAP: $v_1\to v_2$, $l_\text{\tiny CAP}\to l_{\text{\tiny CAP}, t}$, $r_{12}^{ac}\to t_{12}^{ac}$, $a_\text{\tiny CAP}\to a_{\text{\tiny CAP}, t}$, and, correspondingly, $a_{\text{overlap},\alpha}\to a_{\text{overlap},\alpha,t}$. The dependences of the angle $\alpha$ on the angle of the interface inclination $\theta$ in the solutions of Eq.~\eqref{eq:eq22} and Eq.~\eqref{eq:eq26} are different ($\alpha=2\theta \to \alpha=\theta-\theta_t$). Thus, the physical sense of all the characteristic times influencing the dynamics of the transmitted CAP contribution $\left(\frac{\mathrm{d}R}{R}\right)_\text{trans.}$ can be revealed from the analysis accomplished in Secs.~\ref{subsec:free_space} and \ref{subsec:reflection}.

\begin{figure}
    \centering
    \includegraphics{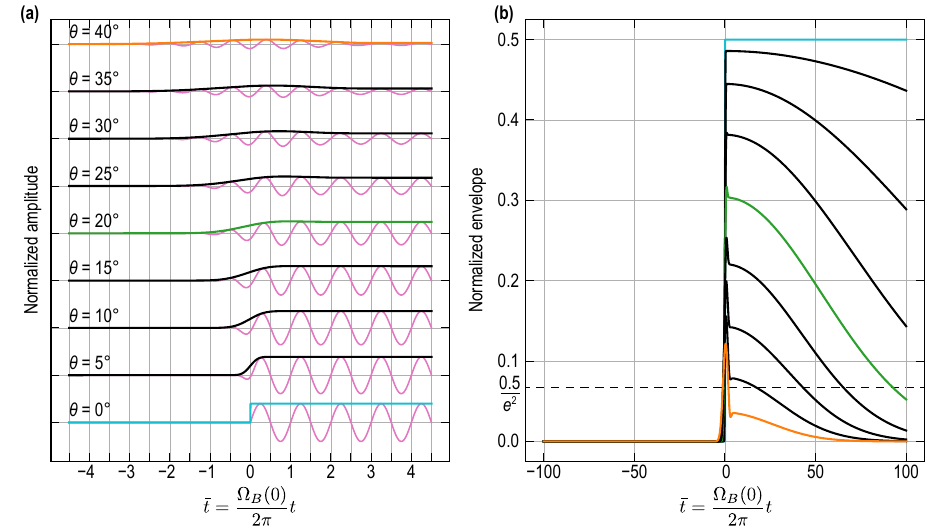}
    \caption{Contribution to the TDBS signal from the transmitted CAP [Eq.~\eqref{eq:eq26}] as a function of the interface inclination angle $\theta$.
    (a) Normalized $\left(\frac{\mathrm{d}R}{R}\right)_\text{trans.}$ (solid oscillating pink curves) for $\theta\in\left[0\si{\degree},40\si{\degree}\right]$ with 5\si{\degree} steps, and associated envelopes (in solid black lines + 3 colored lines). The contributions are upshifted for better visualization. (b) Normalized $\left(\frac{\mathrm{d}R}{R}\right)_\text{trans.}$ envelopes (in solid black lines + 3 colored lines) $\theta\in\left[0\si{\degree},40\si{\degree}\right]$ with 5\si{\degree} steps.} 
    \label{fig:fig12_old_transCAP_contrib}
\end{figure}

Figure~\ref{fig:fig12_old_transCAP_contrib}(a) provides the contribution of the transmitted CAP to the TDBS signal (solid oscillating pink curves) along with their amplitude envelopes, as a function of the inclination $\theta$ of the interface.
In Fig.~\ref{fig:fig12_old_transCAP_contrib}(b), the envelopes of the $\frac{\mathrm{d}R}{R}$ wave packets corresponding to transmitted CAP are presented separately for better appreciation and with a far larger range of values for $\bar{t}$ in order to better appreciate the gradual amplitude decrease of $\left(\frac{\mathrm{d}R}{R}\right)_\text{trans.}$. Three specific envelopes are highlighted with colors, instead of using black lines in Fig.~\ref{fig:fig12_old_transCAP_contrib}, corresponding to the cases $\theta=0$\si{\degree} (blue), $\theta=20$\si{\degree} (green) and $\theta=40$\si{\degree} (orange). By normalizing all spatial variables in Eq.~\eqref{eq:eq26} by the probe beam radius $a_\text{probe}$ and the normalized time $\bar{t}$, it can be easily shown that the expression of $\left(\frac{\mathrm{d}R}{R}\right)_\text{trans.}$ depends on the same three nondimensionalized parameters than in the previous two cases of incident and reflected CAP contributions. These parameters were set to the same values as previously:  $a_\text{\tiny CAP}/a_\text{probe}=1$, $2a_\text{probe}/\lambda_1=5$, $2z_0/\lambda_1=0$. A fourth nondimensionalized parameter arises in the case of the transmission which is the ratio between the acoustic velocity in the two materials, $v_2/v_1$, that was set to 0.75 ($<1$) in Fig.~\ref{fig:fig12_old_transCAP_contrib}. The plotted TDBS signals and envelopes are normalized by $Pt_{12}^{ac}\sqrt{\frac{\pi}{2}}\eta_0l_{\text{\tiny CAP}, t}$

In line with the results of Sec.~\ref{subsec:reflection} for the incident and reflected contributions, in the case of an interface parallel to the free surface of material (1), \textit{i.e.}, with $\theta=0$\si{\degree}, the contribution of the transmitted CAP to the full TDBS signal is starting/emerging when the CAP reaches the interface (blue envelope in Fig.~\ref{fig:fig12_old_transCAP_contrib}). Similar to the contribution from the process of CAP reflection, a characteristic temporal feature associated with the interaction between the interface and the CAP in the process of its transmission across the interface becomes apparent near $\bar{t}=0$ in Fig.~\ref{fig:fig12_old_transCAP_contrib}(b), with its visibility amplified by the inclination angle. Indeed, the contribution of Brillouin oscillations from the transmitted CAP does not diminish as abruptly as the contribution from the reflected CAP (Fig.~\ref{fig:fig10_old_reflCAP_contrib}). Contrarily, the envelope amplitude is decreasing on a larger time scale, \textit{e.g.}, falling to the $1/e^2$ level in more than 90 periods of the Brillouin oscillation $2\pi/\Omega_B(0)$ for $\theta=20$\si{\degree} [green line in Fig.~\ref{fig:fig12_old_transCAP_contrib}(b)], way after the reflected (much less than 1 period) and incident (less than 1 period) contributions. It also means that the transmitted CAP contribution to the TDBS signal is the strongest contribution to the full TDBS signal after few periods of Brillouin oscillations from the instant when the central ray of the delta-CAP reached $z=z_0$. This \enquote{disappearance} time scale can again be related to the evolution of the overlap volume of the CAP and probe light beams. Since the deviation angle $\theta_t$ between the transmitted CAP and the probe light propagation directions is larger compared to the deviation angle $2\theta$ between the reflected CAP and the probe light propagation directions, then the overlap volume is decreasing slower in the case of the transmitted CAP than in the case of the reflected CAP. Note that the smaller the ratio $(v_2/v_1)$, the faster the decrease in amplitude, as the deviation of $\theta_t$ from $\theta$ increases with the diminishing of $(v_2/v_1)$, \textit{i.e.}, the interaction angle $\alpha$ increases with the diminishing of $(v_2/v_1)$.

Note that the previous formula is valid not only for the case where $(v_2/v_1)<1$ but also for the case $1<(v_2/v_1)<1/\sin\theta$. For $(v_2/v_1)>1/\sin\theta$, the CAP incident on the interface exhibits internal reflection and the transmitted CAP becomes evanescent. Hence, the solution in Eq.~\eqref{eq:eq26} is valid for all possible transformations of the CAP in propagating (non-evanescent) transmitted acoustic beam. Under the previous assumption, the transient factor of the transmitted CAP contribution to the TDBS signal hence reads:
\begin{align}\label{eq:eq27}
    G_t\equiv\left[1+\mathrm{erf}\left({\footnotesize\frac{\sqrt{2}x_A(t)}{a_{\text{overlap},\alpha,t}}}+{\footnotesize\frac{\sqrt{2}\sin\alpha v_2t}{a_{\text{\tiny CAP},t}^2\cos^2\alpha}}a_{\text{overlap},\alpha,t}-i\sqrt{\frac{k_1^2}{2}{\footnotesize\frac{a_\text{probe}^2a_{\text{\tiny CAP},t}^2\sin^2\alpha}{a_\text{probe}^2+a_{\text{\tiny CAP},t}^2\cos^2\alpha}}}\right)\right]\,.
\end{align}
Obviously, the dynamics of the factor $G_t$ [Eq.~\eqref{eq:eq27}] can be revealed by scaling the dynamics of $G_r$ in Eq.~\eqref{eq:eq23}.

\subsection{TDBS monitoring of the CAP beam reflection at the interface between optically different materials}
\label{subsec:reflection_with_light_reflection}
\begin{figure}
    \centering
    \includegraphics{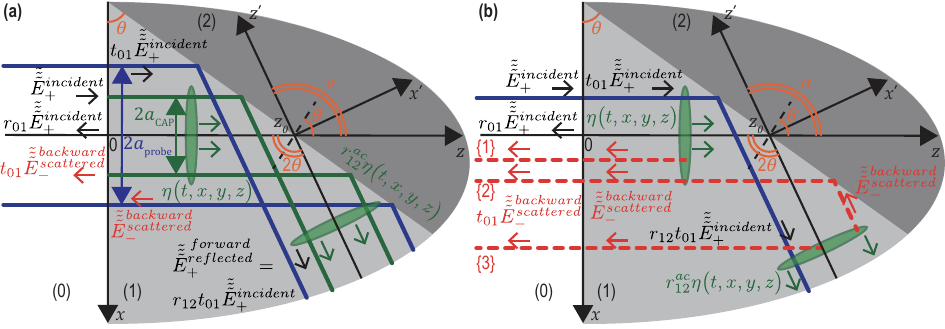}
    \caption{(a)-(b) Sketches of TDBS monitoring of the CAP reflection from the interface between two materials inclined at an angle $\pi/4\le\theta\le\pi/2$ relatively to the free surface ($z=0$) of material (1) ($z>0$) in the case of two optically different materials. The deviation angle, with which the central ray of the coherent acoustic beam deviates from its initial propagation direction, is equal to $2\theta$. (a) The continuous green and blue lines present, inside their Rayleigh ranges, the acoustic and laser beams, respectively. The CAPs are additionally presented by green ellipses to illustrate their strong localization along the propagation direction caused by their ultrashort duration. (b) The probe light path is presented symbolically by a single blue line, while only the CAPs representations by green ellipses are kept. Red dashed rays numbered from \{1\} to \{3\} represent the paths of the light scattered by the incident (\{1\}) and reflected (\{2\} and \{3\}) CAPs in the direction of its heterodyning.}
    \label{fig:fig14_refl_probe_large_angle}
\end{figure}

The theory developed in Sec.~\ref{subsec:reflection} is now extended to the situation where the materials (1) and (2), forming the interface, are in the same geometry as the one presented in Fig.~\ref{fig:fig7_scheme_reflection}, are still optically isotropic, but have different optical and acousto-optical properties. In that case, the reflection of the probe light beam at the interface should be accounted for, and modifications of the initial probe light field in material (1) (where probe light exhibits scattering due to the acousto-optic interaction if a CAP is launched in the sample) are required. Figure~\ref{fig:fig14_refl_probe_large_angle}(a) presents the modification of Fig.~\ref{fig:fig7_scheme_reflection}(a) but with an inclination angle $\pi/4\le\theta\le\pi/2$. The light transmitted in material (2) is not presented, but the probe light beam reflected by the interface is explicitly added. The parameter $r_{12}(n_1,n_2,\theta)$ denotes the probe light reflection coefficient at the interface.

To better sketch the paths of the light scattered by the incident and reflected CAPs in the direction of its heterodyning, Fig.~\ref{fig:fig14_refl_probe_large_angle}(b) is simplified compared to Fig.~\ref{fig:fig14_refl_probe_large_angle}(a). The probe light beam incident on the material interface is now depicted by a single ray, which is shifted up relatively to the central ray, \textit{i.e.}, relatively to the $z$- and $z^\prime$-axes, for clarity. Only the CAPs representations by ellipses are kept. These simplifications allow clear representation of the paths (via red dashed lines numbered from \{1\} to \{3\}) along which the probe light scattered by the incident and reflected CAPs can be detected by heterodyning. Note that these rays are shifted down relatively to the probe light propagation axes, also for clarity. The contributions to the TDBS signal of the two rays directly scattered in the $z$ direction, \textit{i.e.}, rays numbered \{1\} and \{3\} in Fig.~\ref{fig:fig14_refl_probe_large_angle}(b), have been already evaluated in Sec.~\ref{subsec:reflection}. They describe the detection of both incident and reflected CAPs via the scattering of the incident light transmitted from (0) to (1), \textit{i.e.}, via the scattering of $t_{01}\tsup[3]{E}{\vphantom{E}}^\text{incident}_+$. Where $\theta>\pi/4$ and $\alpha=2\theta>\pi/2$ (as now depicted in Fig.~\ref{fig:fig14_refl_probe_large_angle}), the contribution to the full TDBS signal of the light scattered by the reflected CAP would be much smaller than that scattered by the incident CAP. As discussed in Sec.~\ref{subsec:free_space}, in the typical experimental conditions of picosecond ultrasonics ($l_\text{\tiny CAP}\ll a_\text{probe},~a_\text{\tiny CAP}$), the acousto-optic interaction efficiency indeed drops drastically as soon as the interaction is not collinear, because of the exponential factor [Eq.~\eqref{eq:eq19}] driving this efficiency. These two contributions could be potentially comparable only when the inclination angle is approaching $\pi/2$, \textit{i.e.}, in the case where the incident CAP is nearly skimming along the interface. Such a limiting/asymptotic case is of little interest (although the formulas developed in Sec.~\ref{subsec:reflection} are valid) since the configuration depicted in Fig.~\ref{fig:fig14_refl_probe_large_angle} provides an additional way to detect the reflected CAP efficiently compared to the case where probe light was not reflected by the interface (Fig.~\ref{fig:fig7_scheme_reflection}). This efficient way is indeed provided by the 180\si{\degree} backward scattering of part of the probe light [ray \{2\} in Fig.~\ref{fig:fig14_refl_probe_large_angle}(b)] reflected collinearly to the reflected CAP on the material interface, \textit{i.e.}, offering a path with the interaction angle $\alpha=0$\si{\degree}. Note that the inefficient scattering of the reflected probe light, $r_{12}t_{01}\tsup[3]{E}{\vphantom{E}}^\text{incident}_+$ [ray \{3\} in Fig.~\ref{fig:fig14_refl_probe_large_angle}(b)], by the incident CAP is not shown in Fig.~\ref{fig:fig14_refl_probe_large_angle}(b), for clarity. This scattering takes place under the same non-zero interaction angle as the scattering of $t_{01}\tsup[3]{E}{\vphantom{E}}^\text{incident}_+$ [ray \{1\} in Fig.~\ref{fig:fig14_refl_probe_large_angle}(b)] by the reflected CAP and its contribution to the total TDBS signal can be evaluated similarly. 

Starting from here, we would focus only on the dominant contributions to the total TDBS signal due to the scattering at lowest interaction angles between the probe light and the CAP. The mathematical description of the TDBS signal, resulting from the rays due to the probe light backscattered by the reflected CAP to 180\si{\degree} angle, schematically shown in Fig.~\ref{fig:fig14_refl_probe_large_angle}(b) by the red dash-dotted line numbered \{2\}, can be obtained via simple modifications of Eq.~\eqref{eq:eq23}. It is indeed sufficient to set, in Eq.~\eqref{eq:eq23}, $\alpha=0$ and to multiply by the factor $r_{12}^2$ which accounts for the diminishing of this contribution to the TDBS signal caused by the reflections at the interface of the probe and the 180\si{\degree} backward scattered light.

\begin{figure}
    \centering
    \includegraphics{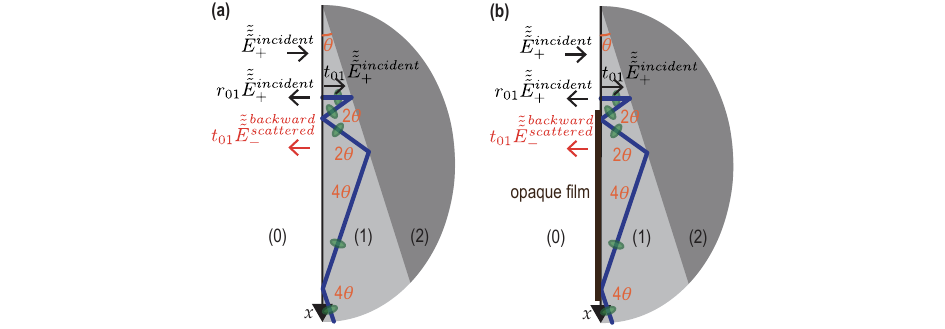}
    \caption{TDBS monitoring of the multiply reflected CAP by the multiply reflected probe light, when the interface between two materials with different elastic and optical properties is inclined at an angle $0<\theta<\pi/4$ relative to the sample surface ($z=0$). The path of the probe light beam is represented by the central ray of the beam, while the launched CAP and the CAP after its subsequent reflections are represented by green ellipses.}
    \label{fig:fig15_refl_probe_small_angle}
\end{figure}

Figure~\ref{fig:fig15_refl_probe_small_angle} provides a simplified scheme of the potential multiple reflections (a single reflection being presented in Fig.~\ref{fig:fig7_scheme_reflection}) of the probe light transmitted from (0) to (1), which could be expected where the interface is inclined at an angle $0<\theta<\pi/4$. In this case, both the CAP and the probe light incident on the (1)/(2) interface, between material (1) and material (2), are reflected and can thus reach back the sample surface where they could undergo a second reflection. As illustrated in Fig.~\ref{fig:fig15_refl_probe_small_angle}(a), this could lead to a trajectory for both CAP and probe light in the form of a zigzag path in material (1).

Potentially, the CAP can be detected along its complete zigzag trajectory (Fig.~\ref{fig:fig15_refl_probe_small_angle}) by the scatterings of the probe light $t_{01}\tsup[3]{E}{\vphantom{E}}^\text{incident}_+$, as it has been described earlier in Fig.~\ref{fig:fig14_refl_probe_large_angle} for the reflected CAP propagating at the interaction angle $\alpha=2\theta$ relative to the $t_{01}\tsup[3]{E}{\vphantom{E}}^\text{incident}_+$ light [see the red dashed ray numbered \{3\} in Fig.~\ref{fig:fig14_refl_probe_large_angle}(b)]. Yet, the angle between the $t_{01}\tsup[3]{E}{\vphantom{E}}^\text{incident}_+$ probe light and the propagating direction of the CAP increases by a $2\theta$ step after each successive reflection of the CAP at the (1)/(2) interface [Fig.~\ref{fig:fig15_refl_probe_small_angle}(a)]. It implies that the TDBS signal would importantly diminish after each reflection of the CAP at the (1)/(2) interface, mostly because of the earlier-discussed exponential amplitude/directivity factor derived in Eq.~\eqref{eq:eq_DB0}. Therefore, it is expected that, along its zigzag trajectory, the CAP could be monitored more efficiently by the probe light that follows the same zigzag trajectory.

After the first reflection at the (1)/(2) interface, the CAP efficiently backscatters the $r_{12}t_{01}\tsup[3]{E}{\vphantom{E}}^\text{incident}_+$ light to 180\si{\degree} angle. While this reflected CAP reaches back the (1)/(0) interface, it reflects again and, while propagating back toward the (1)/(2) interface, efficiently scatters the $r_{10}r_{12}t_{01}\tsup[3]{E}{\vphantom{E}}^\text{incident}_+$ light before the next reflection after which it efficiently scatters the $r_{10}r_{12}^2t_{01}\tsup[3]{E}{\vphantom{E}}^\text{incident}_+$  light, and so on and so forth. Naturally, the intensity of the probe light is progressively diminishing after each reflection and, additionally, the scattered light in its path to the photodetector exhibits an increasing number of reflections, each of them diminishing its amplitude. Additionally, note that the dependence on the angle of incidence of all reflection coefficients is here droped for simplicity while it could diminish even further the amplitudes of all fields. However, it is expected that these quasi-collinear scatterings of probe light would provide larger contributions to the TDBS signal from the reflected CAPs than the scatterings of the $t_{01}\tsup[3]{E}{\vphantom{E}}^\text{incident}_+$ probe light if the inclination angle of the interface is not too small. For example, the ratio of the contributions to the TDBS signal from $t_{01}\tsup[3]{E}{\vphantom{E}}^\text{incident}_+$ and $r_{12}t_{01}\tsup[3]{E}{\vphantom{E}}^\text{incident}_+$ scatterings of the CAP after its first reflection is smaller than \[\frac{\cos 2\theta}{r_{12}^2}\frac{\sqrt{a_\text{probe}^2+a_\text{\tiny CAP}^2}}{\sqrt{a_\text{probe}^2+a_\text{\tiny CAP}^2\cos^2 2\theta}}\exp\left(-\frac{k_1^2}{2}\frac{a_\text{probe}^2a_\text{\tiny CAP}^2\sin^2 2\theta}{a_\text{probe}^2+a_\text{\tiny CAP}^2\cos^2 2\theta}\right).\] Additional diminishing of the evaluated ratio is due to the temporal overlap of the Gaussian envelope and the dynamical $G_r$ function [refer to Eq.~\eqref{eq:eq23}] in the detection by $t_{01}\tsup[3]{E}{\vphantom{E}}^\text{incident}_+$ scattering, which is always smaller than 1 in modulus. In conclusion to this Section, we present in Fig.~\ref{fig:fig15_refl_probe_small_angle}(b) the simplest situation where TDBS involving reflected light and reflected CAP beams provides a potential opportunity for imaging part of the sample which is \enquote{invisible} from the sample surface where the latter is partially covered by an opaque film.

\subsection{TDBS monitoring of the CAP beam transmission across the interface between optically different materials}
\label{subsec:transmission_with_light_reflection}

\begin{figure}
    \centering
    \includegraphics{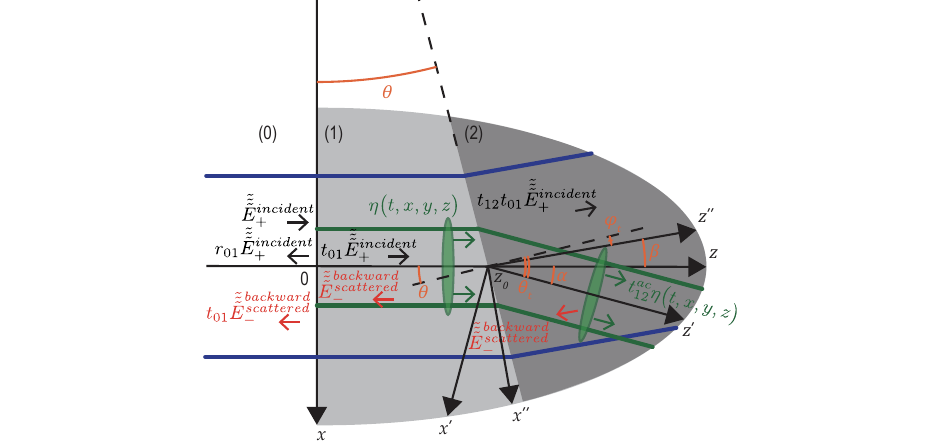}
    \caption{TDBS monitoring of the CAP transmission across the interface between two optically different materials by the probe light beam transmitted across the same interface. $\alpha$ and $\beta$ are the deviation angles with which the central ray of the coherent acoustic beam and that of the probe light beam, respectively, deviate in material (2) from their common initial propagation direction in material (1). Continuous green and blue lines present, inside their Rayleigh ranges, the acoustic and laser beams, respectively. The CAPs are additionally represented by green ellipses to illustrate their strong localization along the propagation direction caused by their ultrashort duration. The angle of sound and light beams on the interface is the same angle $\theta$, while their transmission angles $\theta_t$ and $\varphi_t$, respectively, are different. Note that, for the sake of schematic readability, we have chosen to represent the case where $\beta>0$ and $\alpha<0$. This corresponds to a situation where $n_1<n_2$ and $v_1<v_2$.}
    \label{fig:fig16_trans_probe_small_angle}
\end{figure}

This final Section is dedicated to the extension of the theory developed in Sec.~\ref{subsec:transmission} to a situation analogous to the one presented in Fig.~\ref{fig:fig11_old_scheme_transmission}, the difference being that materials (1) and (2) have different optical properties. Consequently, the probe light beam is refracted when transmitted from material (1) to material (2) thus modifying the initial probe light field in material (2), which exhibits scattering if the CAP launched in the sample is transmitted from material (1) to material (2). The adaptation of Fig.~\ref{fig:fig11_old_scheme_transmission} with an angle $\theta<\pi/4$ is presented in Fig.~\ref{fig:fig16_trans_probe_small_angle} depicting the change in the propagation direction of the probe light upon its transmission from material (1) to material (2). Figure~\ref{fig:fig16_trans_probe_small_angle} introduces the probe light transmission coefficient at the interface, $t_{12}(n_1, n_2, \theta)$, while the angle of the transmitted probe beam direction relative to the $z$-axis is denoted by $\beta=\theta-\varphi_t$, where $\varphi_t$ is the transmission angle of the probe light beam.

The mathematical description of this general case is more cumbersome compared to the previously described cases, but can still be obtained following the previously described approach that lead to Eq.~\eqref{eq:eq26} from a modified version of the integral equation \eqref{eq:eq14}. Few modifications are of course needed. Index 1 should be replaced by 2 where the optical and acousto-optical parameters of material (2) play a role instead of those of material (1). For examples: in the coefficient $P$ [see Eqs.~\eqref{eq:eq12} and \eqref{eq:eq13}] the product of parameters $k_1 n_1^2 p_1$ should now read $k_2 n_2^2 p_2$; in the Gaussian amplitude (directivity) factor, the wavenumber $k_1$ will be changed by $k_2$; etc. Additionally, the TDBS signal diminishes by a factor $t_{12} t_{21}$, which accounts for the fall in the amplitudes of the probe light and the scattered light in their transmission across the (1)/(2) interface. Compared to the previous case treated in Sec.~\ref{subsec:transmission}, the modification of the probe beam cross section in the plane of refraction should be taken into account ($a_\text{probe}\to a_\text{probe}\cos\varphi_t/\cos\theta$), like it was already accounted before for the CAP beam cross section change upon transmission ($a_\text{\tiny CAP}\to a_\text{\tiny CAP,t}=a_\text{\tiny CAP}\cos\theta_t/\cos\theta$). Finally, the inclination angle between the probe light beam and the CAP beam should be evaluated in the second material: $\alpha\to\alpha-\beta=\varphi_t-\theta_t$. This last-mentioned modification indicates that the relative refraction of probe and CAP beams in transmission from (1) to (2) matters, while the individual refractions of these beams from their common initial propagation direction in material (1) are not. So, although it may be slightly more complex, in principle, this more general case is feasible by applying the previously-cited adjustments and carefully changing the spatial variables for integration. Notably, since the probe light beam now propagates along the $z^{\prime\prime}$-axis instead of the $z$-axis, thoughtful considerations are needed. 

Note, that the heterodyning of the acoustically-scattered light by the probe light reflected at (1)/(2) interface and transmitted into (0) is not included in the theory. Potentially, interference pattern of the probe light in the layer (1) should be taken into account in case of its strong reflections at the interfaces (1)/(2) and (1)/(0). However, we leave the detailed description of this situation for the future, when this is required by the experiments. The analysis can be accomplished by the approaches similar to the ones already presented above, but will be just more cumbersome.

\section{Discussion, conclusions and perspectives}
\label{sec:discussion}
The application of the developed theory for particular experimental situations would require, for sure, careful considerations of all the physical assumptions leading to the derived analytical formulas and, correspondingly, some improvements/adjustments of them.

For instance, we have considered for simplicity that a single acousto-optic parameter is sufficient to describe the acousto-optic interaction. This assumption is valid for the case of optically isotropic materials, where s-polarized probe light, \textit{i.e.}, polarized normally to the plane of our figures (which is assumed to be the plane of the probe light incidence and its further propagation), is scattered by the longitudinal CAPs propagating in the same plane. The extensions of the theory to the cases of differently polarized probe light and/or shear instead of compression-dilatation CAP could require the use of up to two photoelastic constants even in optically isotropic materials \cite{hurley_laser_2000}. In optically anisotropic materials, the number of photoelastic constants required for the precise description of the TDBS increases. Moreover, in optically anisotropic materials the incident probe light inside the material should be in general decomposed into light eigenmodes (of ordinary and extraordinary polarizations, for example). The reflection/transmission at the materials interfaces and scattering by the CAPs should be traced for each eigenmode separately, accounting for the possible optical mode conversions upon transmission/reflection by the interfaces \cite{born_principles_1999}, as well as in the photoelastic scattering processes by CAPs \cite{lejman_giant_2014, lejman_ultrafast_2016}. As different optical modes are characterized by different refractive indices, it is expected that the number of different carrier frequencies and of wave packets, simultaneously present in the TDBS signals, would increase in optically anisotropic materials compared to optically isotropic ones \cite{lejman_giant_2014, lejman_ultrafast_2016}. In the presence of materials interfaces, the reflection angle of light in the optically anisotropic material is generally not equal to its angle of incidence, and depends on the crystal axes orientation at the interface, even in the case of reflection without mode conversion. The multimodal probe light is not reflected and refracted along single directions by the material interface between materials (1) and (2). Thereby, different efficiencies of probe light scattering by the CAPs are expected, depending on the closeness between these \enquote{multiple} reflected/refracted light directions and the propagation directions of the reflected/refracted CAPs.

In acoustically (elastically) anisotropic materials, up to three acoustics modes (one quasi-longitudinal and two quasi-transversal modes) can propagate in the same direction at different velocities. The presence of different velocities of CAPs causes the increase in the number of different carrier frequencies \cite{lejman_giant_2014, lejman_ultrafast_2016} and of corresponding wave packets in the TDBS signals. In the presence of material interfaces, the reflection angle of CAP in the elastically anisotropic material is generally not equal to its angle of incidence, and depends on the crystal axes orientation at the interface, even in the case of reflection without mode conversion. The CAPs of different acoustic modes are not reflected and refracted along single directions by the material interface between materials (1) and (2). Thereby, different efficiencies of probe light scattering by the CAPs are expected, depending on the closeness between these \enquote{multiple} reflected/refracted CAPs directions and the propagation directions of the reflected/refracted probe light eigenmodes.

Surely, the dependence of the wave velocities on the wave vector direction in anisotropic crystals would not only have an influence on the TDBS through the multimodality of the wave fields and once the launched CAP and the probe laser beams would have interacted with the material interface. The possible difference in the directions of the phase and group velocities in anisotropic materials could influence the TDBS of probe light beam by the CAP beam even in the absence of material interfaces, \textit{i.e.}, in single homogeneous crystals. When studying crystals, it should be taken into account that the energies carried by the CAP and the probe laser pulse are not along the directions of the phase velocities, which in the geometry considered here are normal to the free surface of the sample, but along the direction of the group velocity directions, that are generally different from the phase velocity directions. Thus, in general, there could be a difference between the group velocities directions of the CAPs and the probe light pulses. This effect influences the overlap volume $l_B$ of the probe light and the CAP beams. For example, in the case when the pump and probe beams are co-focused on the sample surface, the overlap volume between the probe light beam and the CAP would continuously diminish with time from the moment when the photo-generated CAP is completely launched inside the crystal. If the angle between the group velocities of the CAP and the probe light is denoted by $\alpha_g$, then the characteristic distance/depth from the surface where the overlap between the picosecond CAP and the probe light disappears can be estimated as $(a_\text{probe}+a_\text{\tiny CAP})/\tan\alpha_g$. For instance, in an optically isotropic crystal, the direction of the probe light group velocity coincides with the direction of its phase velocity, thus $\alpha_g$ is equal to the angle between the phase and the group velocities of the acoustic eigen modes. To give an example, if the considered material is ceria, which is a cubic material and hence is elastically anisotropic but optically isotropic, the maximum angle $\alpha_g$ for the quasi-longitudinal waves will be 16\si{\degree} and the TDBS signal could disappear when the monitored CAP reaches a distance of $\sim 7$~\si{\micro\meter}, if the characteristic radii of the pump and probe laser beams are about 1~\si{\micro\meter}. Note that, even if the phase and group velocities of the CAP are collinear before the CAP reaches the interface, i.e., if the sample surface is normal to an axis of high symmetry of material (1), they could deviate in their directions after the CAP reflection/transmission by an inclined interface.

The aforementioned comments lead to the expected conclusion that the lower the symmetry of the materials forming the interface, the more challenging the general description of the TDBS signals near the material interfaces could be, although it is still feasible. In the interest of our current research, it is important that the TDBS signal solutions, theoretically developed in Secs.~\ref{sec:TDBS_arbitratry_acField} and \ref{sec:TDBS_arbitratry_angle} for the single-mode probe light and single-mode (longitudinal) CAP, could be used to estimate the magnitudes and durations of all the contributions to the total TDBS signal. These contributions arise from all possible photoelastic interactions between all the CAPs and all the probe light optical beams: they are either directly launched in the sample or generated through reflections/transmissions/mode-conversions. These estimates require the use of the known theories for light and sound in anisotropic materials \cite{born_principles_1999, auld_acoustic_1973} in order to estimate the interaction angles between particular reflected/transmitted CAP pairs/couples and particular probe light reflected/transmitted modes, as well as to evaluate the velocities of the interacting modes. We have not discussed/analyzed in details the particular TDBS situations of probe light and/or CAP reflection/transmission, \textit{e.g.}, total reflection where evanescent waves are involved. However, the already derived formulas could be extended to deeper analysis, when required. Stated differently, we have avoided to discuss in the text and to represent in the figures the experimental situations when the inclination angle on the interface relative to the front surface of the sample is approaching 90\si{\degree} (material interfaces nearly normal to the sample surface). It is worth mentioning that in this particular geometry it could be required to take into account the contributions to the TDBS signal from the probe light scattered by the CAPs generated directly in the material (2). Indeed, we recall that in all the figures in Sec.~\ref{sec:TDBS_arbitratry_angle}, even in Fig.~\ref{fig:fig14_refl_probe_large_angle} presenting the case of large inclination angles, $\pi/4\le\theta<\pi/2$, the intersection of the materials interface with the sample surface was assumed to be sufficiently far from the acoustic beam core depictured by the cylinder with radius equal to the CAP beam radius. Thus, the eventual CAP launched by the pump laser radiation directly in material (2) when material (2) constitutes a part of the sample surface and is either absorbing light itself or have a deposited absorber on the top, was completely neglected because of negligible optical energy in the edges/extremity of the Gaussian pump laser focus. It was assumed in Sec.~\ref{sec:TDBS_arbitratry_angle} that the CAP could be launched in material (2) only through the interface between materials (1) and (2).

However, real experiments impose some conditions on the observations of the TDBS near the interfaces. For example, the CAP and the probe light launched from the surface of material (1) should not be significantly attenuated (due to the absorption and/or diffraction), and additionally the distance of the interface from the sample surface should not exceed the coherence length of the probe light in material (1) \cite{lin_phonon_1991, sandeep_3d_2021}. The coherence length is a natural limitation, while the theory developed in this manuscript explicitly takes into account that the probe laser radiation is not continuous but is a periodic sequence of femtosecond laser pulses. It corresponds to the maximal distance, between the moving CAP and the stationary surfaces of the sample, that provides the temporary overlap on the photodetector between the moving CAP-scattered light and the fixed/stationary reflected light pulses needed for the heterodyning. Therefore, if the interface is too deep from the surface, \textit{i.e.}, at distances exceeding the minimum of absorption, diffraction and coherence lengths, then it would be necessary, for probing the interface with TDBS, to move (along the sample surface) the pump and probe laser foci closer to the intersection of the interface with the sample surface. The amplitude of the CAPs generated directly in material (2) will increase respectively. In fact, it would be progressively more and more difficult to neglect the contributions to the full TDBS signal from the CAPs directly launched into material (2), especially when the inclination angle between the interface and the sample surface is approaching 90\si{\degree}.

In general, the CAPs photo-generated in both materials should be taken into account if the pump and probe laser beams significantly overlap with both materials on the sample surface. For instance, an equivalent situation is obtained when the three-dimensional TDBS imaging is performed by scanning the position of the co-focused pump and probe light on the sample surface \cite{sandeep_3d_2021}. In such cases where the laser foci overlap with the intersection of the interface and the sample surface, at least two different dominant carrier frequencies are expected in the TDBS signal. They correspond to the TDBS in the two materials separately, even without accounting for the reflections/transmissions of the launched CAPs by the interface. Recently, it was demonstrated that, via revealing two frequencies, TDBS signals near the sample surface can be used for mapping the inter-grain boundaries on the surface of a polycrystalline sample \cite{sandeep_3d_2021}. However, a detailed analysis of this possible situation is beyond the scope of our current discussion.

In relation to the previously-discussed role of the finite duration of the probe laser pulses on the TDBS depth of imaging linked to the coherence length, it is worth mentioning here that, theoretically, the influence of the laser pulses duration on the carrier frequencies and the durations of the TDBS wave packets is also expected. To increase the coherence length, the probe laser pulses duration could be increased from femtoseconds and sub-picoseconds to picoseconds. These effects could be included in the extensions of the developed theory when required.

An important point in the applications of the developed theory for the interpretation of the experiments would be the verification of its validity in each particular experimental geometry. This is required, since both acoustical and optical reflection/transmission coefficients depend on the angle of light/sound incidence on the material interface. In general, these dependencies modify the plane wave spectra of both waves and could cause the broadening of the directivity patterns of light and sound beams, diminishing their respective Rayleigh ranges. In general, the different rays composing light and sound cones, shown in Fig.~\ref{fig:fig6_ray_cones}, are transmitted/reflected by interfaces with different coefficients and it is decreasingly precise to approximate the beam transmission/reflection by a single coefficient, when the dependencies of the coefficients on the angle of incidence become steeper and steeper. These circumstances should be also taken into account when extending the theory to curved interfaces, like in the cases of CAP reflections from spheres or cylinders.

It is also worth mentioning that the theory developed in Sec.~\ref{sec:TDBS_arbitratry_angle} for unipolar CAPs corresponds well to the experimental conditions at high pressure in a diamond anvil cell, where unipolar CAPs are launched from the laser-irradiated surface of a metallic optoacoustic transducer within the transparent polycrystalline material loading it \cite{nikitin_revealing_2015, kuriakose_picosecond_2016}. However, in the case of a mechanically free sample surface, only the destructive ablation regime of optoacoustic conversion launches nearly unipolar CAPs in the sample, while the nondestructive thermo-elastic optoacoustic conversion launches bipolar pulses. For the latter case, the developed theory should be modified correspondingly. Keeping the description of all the spatial and temporal profiles by Gaussian functions, it is natural to model the bipolar strain CAP profile by the first derivative of the Gaussian function. Then, the necessary integrals to evaluate in the theory could be, depending on the convenience, either immediately transformed to the same type of integrals earlier done via the integration by parts, or, differently, the integrals of Gaussian-derivatives-type CAP profile could be transformed into the derivatives of the previous integrals with unipolar strain pulses. In any case, the final results for bipolar CAPs could be described analytically in terms of the same functions as in the case of unipolar CAPs, \textit{i.e.}, in terms of the trigonometric, Gaussian and Error functions.

The theory could be also extended to the case where the probe light is incident at an angle on the sample and where the probe light transmitted in the sample is efficiently scattered by the CAP in the direction of its heterodyning, determined by the direction of the probe light reflection from the sample surface \cite{lomonosov_nanoscale_2012, cote_refractive_2005, tomoda_tomographic_2007}. It could be extended also to the cases where the direction of the scattered light used for heterodyning could be chosen/controlled independently of the probe light incidence direction \cite{ocelic_pseudoheterodyne_2006}. Finally, it is worth recalling here that the current theory takes into account only the processes of probe light backscattering. The CAP beam can be potentially probed by forward Brillouin scattering, if the experimental geometry is arranged for heterodyning the forward scattered light or its reflections at the sample surfaces/interfaces \cite{matsuda_time-domain_2018}.

In conclusion, we have developed a straightforward theoretical analytical description for time-domain Brillouin scattering (TDBS) when probe light and coherent acoustic pulse (CAP) Gaussian beams propagate at an angle to each other. We have demonstrated its potential applications by predicting modifications in TDBS signals when acoustic pulses and probe light interact with an inclined interface between two materials. Our theory reveals that the transient TDBS signals, accompanying CAP reflection from or transmission through the interface, strongly depend on the inclination angle of the interface. Therefore, these signals can be utilized to assess the inclination of an interface through a single local TDBS measurement. This perspective has been recently validated through experiments investigating the destruction of a single crystal plate under non-hydrostatic pressure loading in a diamond anvil cell \cite{sandeep2023}.

\section*{Funding}
The research leading to these results received funding from French National Research Agency (ANR, France) under Grant Agreement No ANR-18-CE42-0017. This work was also supported by the R{\'e}gion Pays de la Loire through the RFI Le Mans Acoustique (project title \enquote{Paris Scientifique OPACOP 2018}, salary of T.T. when he was a PhD student).

\section*{CC-BY public copyright license}
This research was funded, in whole or in part, by the French National Research Agency (ANR, France) under Grant Agreement No ANR-18-CE42-0017. A CC-BY public copyright license has been applied by the authors to the present document and will be applied to all subsequent versions up to the Author Accepted Manuscript arising from this submission, in accordance with the grant’s open access conditions.



 \bibliographystyle{elsarticle-num} 
 \bibliography{TDBS_theory_non_collinear_beams}





\end{document}